\documentclass[twocolumn]{aastex62}

\hypersetup{linkcolor=red,citecolor=cyan,filecolor=cyan}

\newcommand{\msun}{\mbox{$\rm M_\odot$}}

\newcommand{\replace}[1]{\textcolor{black}{#1}}

\newcommand{\insitu}{\textit{in situ}}
\newcommand{\exsitu}{\textit{ex situ}}
\newcommand{\qsay}[1]{``#1''}
\newcommand{\bracket}[1]{\left\langle#1\right\rangle}
\newcommand{\fyoungdisk}{$f_{\rm disk}^{\rm 100\,Myr}$}
\newcommand{\fbirthdisk}{$f_{\rm disk}^{\rm at\, birth}$}

\graphicspath{{./}}

\shorttitle{New Horizon: disk and spheroid formation}

\usepackage{natbib}
\usepackage{lipsum}
\usepackage{amsmath}
\usepackage{comment}
\usepackage{multirow}
\usepackage{enumitem}
\usepackage[utf8]{inputenc}
\usepackage[flushleft]{threeparttable}

\begin{document}
\title{New Horizon: On the origin of the stellar disk and spheroid of field galaxies at \replace{$z=0.7$}}

\author{Min-Jung Park}
\affil{Department of Astronomy and Yonsei University Observatory, Yonsei University, Seoul 03722, Republic of Korea}

\author{Sukyoung K. Yi}
\affil{Department of Astronomy and Yonsei University Observatory, Yonsei University, Seoul 03722, Republic of Korea}

\author{Yohan Dubois}
\affil{Institut d’Astrophysique de Paris, Sorbonne Universités, UMPC Univ Paris 06 et CNRS, UMP 7095, 98 bis bd Arago, 75014 Paris, France}

\author{Christophe Pichon}
\affil{Institut d’Astrophysique de Paris, Sorbonne Universités, UMPC Univ Paris 06 et CNRS, UMP 7095, 98 bis bd Arago, 75014 Paris, France}
\affil{Korea Institute of Advanced Studies (KIAS) 85 Hoegiro, Dongdaemun-gu, Seoul, 02455, Republic of Korea}

\author{Taysun Kimm}
\affil{Department of Astronomy and Yonsei University Observatory, Yonsei University, Seoul 03722, Republic of Korea}

\author{Julien Devriendt}
\affil{Dept of Physics, University of Oxford, Keble Road, Oxford OX1 3RH, UK}

\author{Hoseung Choi}
\affil{Institute of Theoretical Astrophysics, University of Oslo, Postboks 1029, Blindern, 0315 Oslo, Norway}

\author{Marta Volonteri}
\affil{Institut d’Astrophysique de Paris, Sorbonne Universités, UMPC Univ Paris 06 et CNRS, UMP 7095, 98 bis bd Arago, 75014 Paris, France}

\author{Sugata Kaviraj}
\affil{Centre for Astrophysics Research, School of Physics, Astronomy and Mathematics, University of Hertfordshire, College Lane, Hatfield AL10 9AB, UK}

\author{Sebastien Peirani}
\affil{Observatoire de la Côte d’Azur, CNRS, Laboratoire Lagrange, Bd de l’Observatoire,Université Côte d’Azur, CS 34229, 06304 Nice Cedex 4, France}

\begin{abstract}
The origin of the disk and spheroid of galaxies has been a key open question in understanding their morphology. 
Using the high-resolution cosmological simulation, New Horizon, we explore kinematically decomposed disk and spheroidal components of 144 field galaxies with masses greater than $10^9\,\msun$ at $z=0.7$.
The origins of stellar particles are classified according to their birthplace (\insitu\, or \exsitu) and their orbits at birth.
Before disk settling, stars form mainly through chaotic mergers between proto-galaxies and become part of the spheroidal component.
When disk settling starts, we find that more massive galaxies begin to form disk stars from earlier epochs; massive galaxies commence to develop their disks at $z\sim1-2$, while low-mass galaxies do after $z\sim1$. 
The formation of disks is affected by accretion as well, as mergers can trigger gas turbulence or induce misaligned gas infall that prevents galaxies from forming co-rotating disk stars.
The importance of accreted stars is greater in more massive galaxies, especially in developing massive spheroids.
A significant fraction of the spheroids comes from the disk stars that are perturbed, which becomes more important at lower redshifts.
Some ($\sim12.5\%$) of our massive galaxies develop counter-rotating disks from the gas infall misaligned with the existing disk plane, which can last for more than a Gyr until they become the dominant component, and flip the angular momentum of the galaxy in the opposite direction.
The final disk-to-total ratio of a galaxy needs to be understood in relation to its stellar mass and accretion history. 
We quantify the significance of the stars with different origins and provide them as guiding values. 
\end{abstract}


\section{Introduction} 
\label{section1} 

In the local Universe, galaxies have a wide variety of morphology ranging from disk-dominated spiral galaxies to bulge-dominated elliptical galaxies \citep{Hubble1926EXTRA-GALACTICHUBBLE}.
From many observational results, it is also well-established that the morphology of galaxies is highly correlated with other properties, such as luminosity (mass), color, and star formation rate \citep[e.g.,][]{Conselice2006TheSystem,Driver2006ThePlane,Benson2007LuminosityFunction,Ilbert2010GalaxySurvey}.
The \replace{two most distinctive} components of a galaxy behind this morphological diversity are supposedly their disk and spheroidal components, and their distinct stellar populations and kinematics suggest that they were formed through different formation mechanisms.
Therefore, identifying the origin of these structures is an important step towards understanding the formation of galaxies with different shapes.

The conventional view of disk formation is that stellar disks are formed when shock-heated gas slowly cools down and collapses into dark matter (DM) haloes while conserving its angular momentum
\citep{FallS.MichaelEfstathiou1980FormationHaloes,Mo1998TheDiscs,Cole2000HierarchicalFormation}.
Recent hydrodynamic simulations have further elaborated on this classical understanding by showing that gas can quickly collapse and form filamentary structures before it reaches galactic haloes, the so-called \qsay{cold-mode} accretion \citep{Keres2005HowGas,Ocvirk2008BimodalSimulation,Dekel2009FormationSpheroids}. 
This cold stream through the cosmic filaments is thought to be a dominant process of gas accretion for low-mass or high-redshift galaxies.
In addition, as gas accumulates angular momentum as it travels along the filament, later-infalling gas would carry higher angular momentum into a galaxy \citep{Pichon2011RiggingDiscs, Kimm2011TheRevisited, Stewart2011OrbitingAccretion}. 
\replace{Continuous supply of gas seems to play an important role in the formation of disk galaxies; for example, gas rich major mergers are proposed as one possible channel for the formation of massive disk galaxies \citep[e.g.,][]{Barnes2002FormationGalaxies,Governato2009FormingMerger}.}

In a hierarchical Universe, galaxies grow into massive ellipticals via mergers. 
As a result of mergers, galaxies develop dispersion-dominated spheroidal components as the aligned orbits of disk stars are disturbed \citep{Toomre1977MergersConsequences, Negroponte1983SimulationsGalaxies}.
Therefore, disk-dominated galaxies are generally expected to have experienced less violent events that could destroy their disks, and the formation of massive elliptical galaxies is believed to be the result of numerous hierarchical merging.
Accordingly, merger history has been considered as the most important factor in determining the morphology of galaxies, as shown in the results of several numerical studies \citep[e.g.,][]{Scannapieco2009TheUniverse, Martig2012AGalaxies, Aumer2014TheGalaxies, Martin2018TheTime}.

``Accretion'' of the stars of disrupted satellite galaxies during the process of mergers is also an important channel to the development of spheroids.
These accreted stars are believed to contribute to the growth of the outskirts of galaxies \citep{Oser2010TheFormation}, especially to the formation of the halo components \citep{Searle1978CompositionsHalo, Zolotov2009TheHalos, Tissera2012ChemicalGalaxies, Tissera2013StellarProperties, Cooper2015FormationGalaxies}.
From a hierarchical point of view, more massive galaxies are expected to have more accretion \citep{Rodriguez-Gomez2016TheStars,Lee2017FormationMass}, and some studies using cosmological large-volume simulations have argued that accretion plays a major role in driving the morphology of the massive galaxies to be spheroidal \citep{Dubois2016TheFeedback, Clauwens2018TheFormation,Martin2018TheTime}. 
For example, \cite{Dubois2016TheFeedback} have shown that the morphology of galaxies at fixed mass strongly depends on cosmic accretion.

Spheroids can grow through several other processes besides mergers.
Many numerical studies have shown that turbulent gas-rich disks at high redshifts tend to fragment into massive clumps by gravitational instabilities;
these clumps formed \insitu\, in the proto-disks can migrate to the central regions and coalesce to form a central bulge \citep{Noguchi1999EarlyDisks,Elmegreen2008BulgeGalaxies,Dekel2009FormationSpheroids}.
Another mechanism that contributes to the growth of spheroids is related to the formation of bars; a bar can induce gas inflow into the central regions, enhancing central star formation to form disk-like bulges \citep{Athanassoula2005OnSimulation}.

Misaligned gas infall is also thought to contribute to disk shrinkage and redistribution of mass from disks to spheroids \citep{Scannapieco2009TheUniverse,Zolotov2015CompactionNuggets}.
Occasionally, a counter-rotating component develops in a galaxy during the misaligned infall, which can cause reorientation of disks. 
\citet{Aumer2013IdealizedHaloes} found in their simulation that disk fraction decreases as galaxies have experienced more frequent reorientation.
Several studies have claimed that the {\em spin alignment} of gas accretion is more crucial in determining the final morphology of galaxies than the {\em frequency} of mergers \citep{Pichon2011RiggingDiscs, Sales2012TheGalaxies}.
All of the above-mentioned internal and external processes affect the formation of spheroids, but it is still unclear which process is the most significant.

\begin{figure}
    \linespread{1.0}\selectfont{}
    \includegraphics[width=\columnwidth]{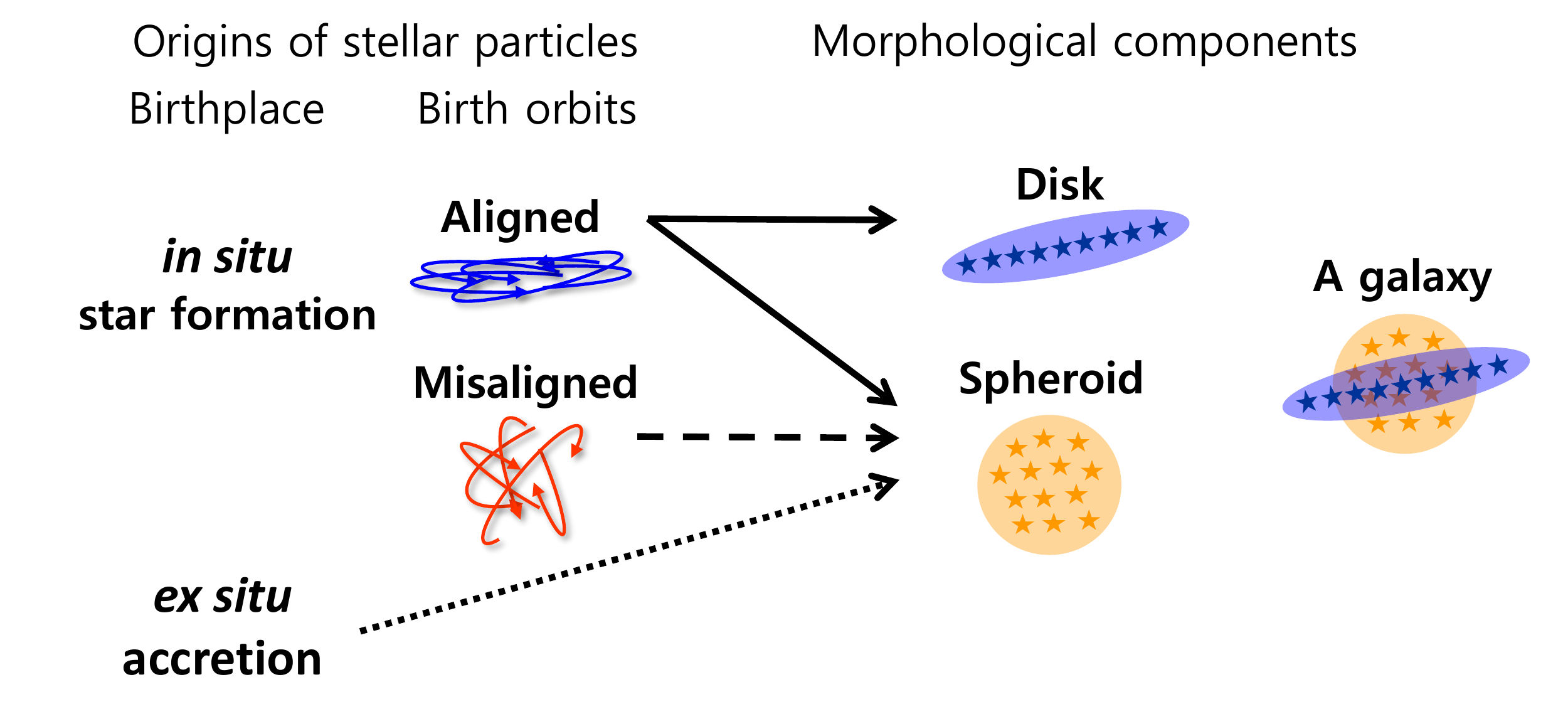}    
    \caption{A schematic diagram summarizing the possible channels to the disk and spheroidal components of a galaxy, based on the origins of stellar particles.} 
    \label{fig:intro}
\end{figure}

Hence, the origin of disk and spheroid may be speculated as follows.
Spheroids form from low angular momentum material in the early stage of galaxy formation where chaotic mergers between proto-galaxies are frequent.
Galactic disks develop later with \insitu\, star formation from coherent accretion \replace{of gas.} 
As galaxies evolve, new spheroidal components can further grow from disk instability, misaligned gas infall, and hierarchical merging.
Therefore, as the schematic diagram of Figure~\ref{fig:intro} shows, disks are mainly composed of the stars formed \insitu\ with co-rotating co-planar initial orbits (``aligned'' orbits), while spheroids consist of the stars formed \insitu\ with non co-planar initial orbits (``misaligned'' orbits, dashed arrow), as well as the stars with disk-origins (solid arrow), and the stars formed \exsitu\ and later accreted (dotted arrow).

In this study, we use the New Horizon simulation \replace{(Dubois et al. in prep.)}, a high-resolution cosmological zoom-in simulation that includes a statistically significant number of galaxies, to understand the origin of disk and spheroidal components.
This study aims to quantify the relative importance of the channels to the disk and spheroidal components (see Figure~\ref{fig:intro}), as probed by our sample.
Specifically, this study attempts to answer the following questions:
1) when do galaxies start to predominantly form disk stars? 
2) what is the contribution of each channel to the formation of spheroids?
and 3) how does galactic morphology evolve when a galaxy develops a counter-rotating component from gas infall misaligned with the existing disk plane?

This paper is organized as follows. In Section~\ref{section2}, we describe the New Horizon simulation, galaxy identification, and sample selection. 
Section~\ref{section3} presents the kinematic decomposition techniques used to identify the disk and spheroidal components and to measure the mass ratio of the components as a morphological indicator.
In Section~\ref{section4}, we explore the formation of disk and spheroidal components of the New Horizon galaxies by tracing the origin of stellar particles.
In Section~\ref{section5}, we examine the evolution of kinematic morphology until $z=0.7$ and provide the estimates of the contribution from different channel to the disk and spheroidal components.
Finally, we summarize our results in Section~\ref{section6}.

\section{Methodology} 
\label{section2} 

\subsection{The New Horizon simulation} \label{sec:2.1}
We use the New Horizon simulation (Dubois et al. in prep.), a high-resolution cosmological simulation performed with an Eulerian hydrodynamics code with adaptive mesh refinement (AMR), {\sc ramses} \citep{Teyssier2002CosmologicalRefinement}. 
\replace{The parent simulation, Horizon-AGN \citep{Dubois2014DancingWeb} is a large volume  simulation (with box size of $\rm 100\,Mpc/h$) that successfully reproduces galaxies in the various cosmic environments including 14 galaxy clusters with $M_{vir}>10^{14}\,\msun$. 
While the maximum spatial resolution of $\rm \sim1\,kpc$ (with a dark matter mass resolution of $8\times10^7\,\msun$ and a stellar mass resolution of $2\times10^6\,\msun$) allows us to understand the general evolutionary trends of a large number of galaxies ($\sim$85,000 galaxies above $10^9\,\msun$ at $z=0$), it is still not sufficient to study the detailed structures of galaxies.}

\begin{figure}
    \linespread{1.0}\selectfont{}
    \includegraphics[width=\columnwidth]{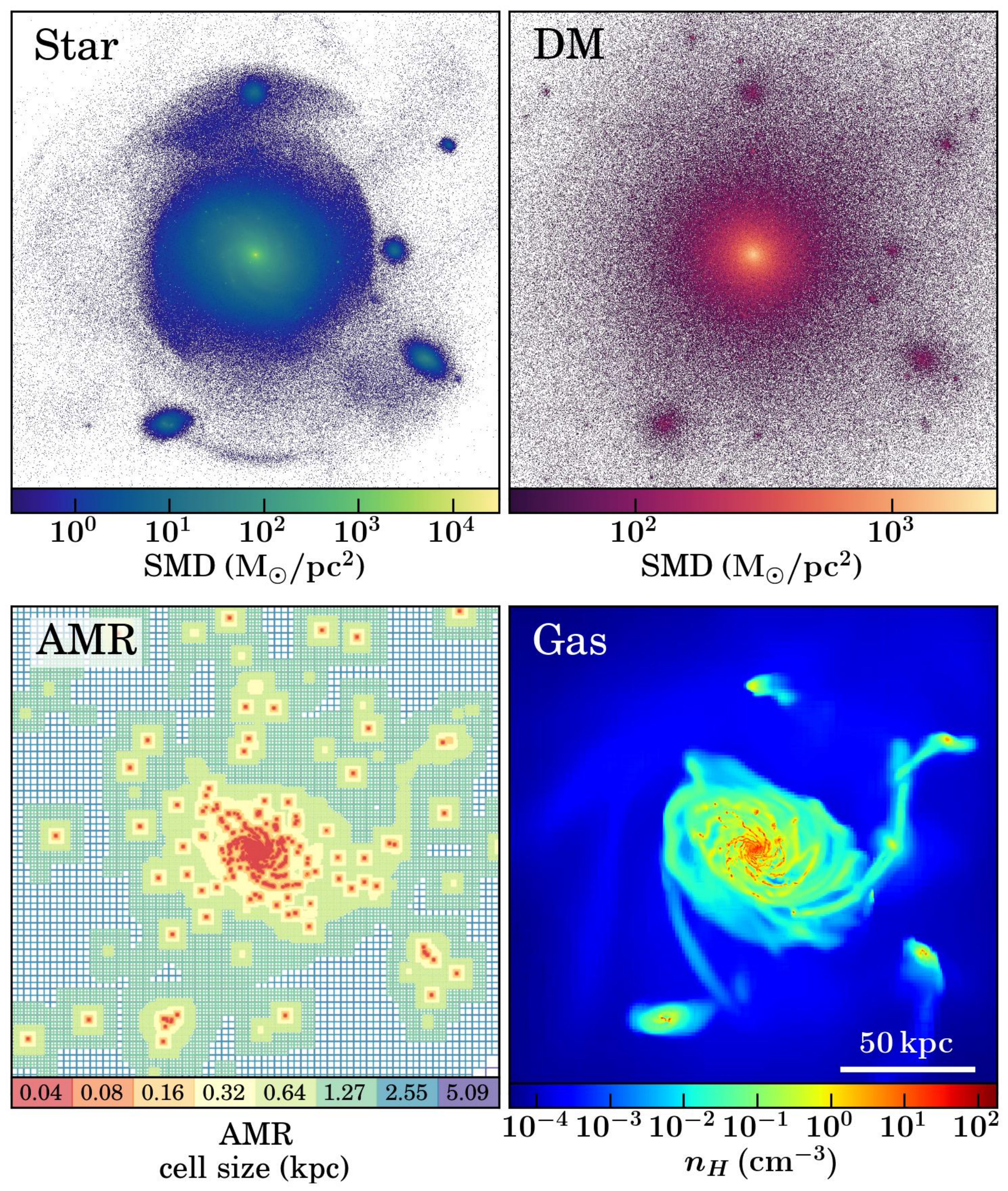}    
    \caption{A projected map of the stellar particles, dark matter particles, AMR grid, and gas density of a galaxy at $z=0.7$ in the New Horizon simulation. 
    Note that the minimum mass resolutions for the dark matter and stellar particles are  \replace{$M_{\rm DM,res}\sim10^6\,\msun$} and \replace{$M_{\rm *,res}\sim10^4\,\msun$}, respectively. 
    The maximum spatial resolution is $\rm 40\,pc$ (physical scale).} 
    \label{fig:allmap}
\end{figure}

\replace{The New Horizon simulation zooms in on a sphere with a radius of $\rm 10\,Mpc$ (comoving) in a ``field'' environment}\footnote{The mass of the most massive halo in the New Horizon volume at $z=0.7$ is $\sim3.9\times10^{12}\,\msun$. 
In order to quantify the environment, we counted the 10 nearest neighbors (with masses greater than $10^9\,\msun$) for each of our sample and divided by the circular enclosed area. 
Based on this quantitative term, galaxy population can be separated as follows: field galaxies ($\Sigma<10\,\rm Mpc^{-2}$), group galaxies ($\Sigma\sim100\,\rm Mpc^{-2}$), and cluster galaxies ($\Sigma\sim1000\,\rm Mpc^{-2}$) (e.g., \citealt{Smith2005EvolutionGalaxies}).
Given that the median value of the projected surface density for our sample is $\Sigma\sim5.6\,\rm Mpc^{-2}$ (at $z=0.7$, using 100 random projections), most of our sample seems to be biased toward field galaxies.}, 
\replace{that is extracted from the Horizon-AGN simulation with a dark matter mass resolution of $10^6\, \msun$ and a minimum stellar mass resolution of $10^4\, \msun$. 
The simulation has reached $z=0.7$ with maximum spatial resolution of $\rm 40 \, {\rm pc}$ (physical scale, at $z=0.7$)\footnote{We are trying to find more computing time to reach $z=0$.}.} 
Note that this is the highest spatial resolution \replace{simulation yet in a cosmological volume.}
Figure~\ref{fig:allmap} shows the maps of the stellar, dark matter, and gas distributions and the AMR grid of a sample galaxy in the New Horizon simulation. 
The simulation adopts cosmological parameters consistent with the WMAP-7 data \citep{Komatsu2011SEVEN-YEARINTERPRETATION}:
Hubble constant $H_{0}=70.4\,{\rm km\,s^{-1}\,Mpc^{-1}}$, 
total mass density $\Omega_{m}=0.272$,
total baryon density $\Omega_{b}=0.0455$, 
dark energy density $\Omega_{\Lambda}=0.728$, 
amplitude of power spectrum $\sigma_{8}=0.809$, 
and power spectral index $n_{s}=0.967$.

Radiative cooling is modeled following \cite{Sutherland1993CoolingPlasmas} and \cite{Rosen1995GlobalGalaxies}, which allows gas to cool to $1\,{\rm K}$ through primordial and metal cooling.
Gas can also be heated via a uniform ultraviolet radiation after the reionization epoch at $z=10$ based on \citet{Haardt1996RadiativeBackground}.
Stars can form out of gas cells with a hydrogen number density greater than $n_H=10\,{\rm cm^{-3}}$ and a temperature lower than $2\times10^4 \,{\rm K}$, following the Schmidt law \citep{Schmidt1959TheFormation}. 
Instead of adopting a fixed star formation efficiency per free-fall time ($\epsilon_{\rm ff}$), we determine the local $\epsilon_{\rm ff}$ based on the local thermo-turbulent condition \citep{Kimm2017Feedback-regulatedReionisation}.
When each stellar particle becomes older than $5\,{\rm Myr}$, we assume a supernovae explosion returning 31\% of the stellar mass to the surroundings. 
Supernova feedback is modeled using a mechanical feedback scheme \citep{Kimm2014EscapeStars}. 
We adopt the Chabrier initial mass function \citep{Chabrier2005The2005} with upper and lower mass cutoffs of $0.1\,\msun$ and $150\,\msun$, respectively.

Black holes, implemented as sink particles, form in cells where both the gas and stellar densities are above the threshold of star formation, with a seed mass of $10^4\,\msun$.
The black holes grow on the basis of Bondi--Hoyle--Lyttleton accretion \citep{Hoyle1939TheVariation,Bondi1944OnStars}, and the maximum accretion rate is limited to the Eddington rate. 
AGN feedback is modeled in two different ways depending on the ratio of the gas accretion rate to the Eddington limit by following \citet{Dubois2012Self-regulatedSimulations}.
If the gas accretion rate is lower than $1\%$ of the Eddington rate, the black hole releases mass, momentum, and energy in the form of jets (radio mode). \replace{The jet efficiency depends} on the spin of the black hole that is evolved with the model of \cite{Dubois2014BlackSimulations}, with a spin up rate and an efficiency for the jet mode following magnetically arrested disks simulations \citep{Mckinney2012GeneralHoles}.
Conversely, if the gas accretion rate is higher than $1\%$ of the Eddington rate, the black hole deposits thermal energy isotropically (quasar mode).
A more detailed description of the simulation can be found in Dubois et al. (in prep.).

\subsection{Galaxy and halo identification and sample selection} \label{sec:2.2}
Simulated galaxies are identified using the {\sc AdaptaHOP} algorithm \citep{Aubert2004TheHaloes} with the most massive sub-node mode \citep{Tweed2009BuildingSimulations} applied for stellar particles. 
At least 50 stellar particles are required to be identified as a galaxy, and the position of the stellar particle with the highest spatial density is considered to be the center of the galaxy. 
Dark matter haloes are also identified using the {\sc AdaptaHOP} algorithm.
Since New Horizon is a zoom-in simulation of Horizon-AGN, galaxies close to the boundary of the zoom-in region could be polluted with low-resolution (more massive) DM particles that are initially located outside the zoom-in regions. 
Thus, we limit our analysis to galaxies more massive than $10^9\,\msun$ in haloes with a contaminated fraction of less than \replace{4\%}\footnote{If a galaxy is in a subhalo, both sub and host halo should have low-resolution particles with number fraction lower than 4\% to be part of the sample.}.

Due to the nature of {\sc AdaptaHOP}, star-forming clumps inside a galaxy are identified as different galaxies; stars in these local density peaks are not taken into account as members of the main galaxy in question.
To avoid such subtraction, we include all the substructures (star-forming clumps or dwarf galaxies) inside $R_{90}$ of the galaxy and re-measure  $R_{90}$ to make sure that it contains 90\% of the total stellar mass.
Some of the galaxies in the sample are in the process of merging with massive companions.
In such cases, their morphologies are likely to be highly disturbed, which makes their disk and spheroid structures unreliable.
Therefore, we exclude the galaxies from the sample if they have satellite galaxies, within their $R_{90}$, with masses higher than 10\% of the stellar masses of the host galaxies.
The satellite galaxies in these cases are excluded from the analysis as well.
Finally, we limit our sample to the galaxies with merger trees available up to at least $z=4$.

The resulting main sample consists of 144 galaxies with masses greater than $10^9\,\msun$ (24 \replace{massive} galaxies with $M_{\rm stellar} \geq 10^{10}\,\msun$ and 120 \replace{low-mass} galaxies with $10^{9} \leq M_{\rm stellar} < 10^{10}\,\msun$.) at $z=0.7$, where our simulation stops. 
\replace{Of the 144 selected sample, 91 galaxies (13 massive galaxies and 78 low-mass galaxies) are contamination free and 131 galaxies (20 massive galaxies and 111 low-mass galaxies) have contaminated fraction below 1\%. The properties of these galaxies (e.g., galaxy mass function, halo mass function, star formation rate density, etc.) are in reasonable agreement with observations at the corresponding redshift considering the small sample size, which will be presented in detail in the forthcoming paper that properly introduces the New Horizon project (Dubois et al. in prep.).}

\section{Kinematic decomposition}
\label{section3}

\subsection{Definition}
\label{sec:3.1}

Disks and spheroids are thought to be the most distinct components of a galaxy in terms of kinematics and stellar populations. 
In order to investigate the formation and evolution of each component, it is first necessary to decompose galaxies into these two components.
We use the circularity parameter \citep{Abadi2003SimulationsDisks} to carry out the kinematic decomposition by quantifying the orbital property of each stellar particle in a galaxy.
First, we define the galactic rotational axis (spin axis) as the direction of the net angular momentum vector using the stars within $R_{90}$. 
The circularity parameter is calculated based on the specific angular momentum of each stellar particle on the galactic spin axis ($J_{z}$), normalized to that of an expected circular orbit with the same energy as the stellar particle ($J_{cir}(E)$): 
\begin{equation}
\label{eq:circularity}
\epsilon=J_{z}/J_{cir}(E).
\end{equation}
By definition, a star on a circular orbit in the galactic plane would have a circularity parameter of 1, while the circularity parameters of a group of randomly-orbiting stars would be distributed centered around 0.

We adopt a fixed cutoff of $\epsilon=0.5$ to determine whether a stellar particle is in the ordered (disk) or disordered (spheroidal) component.
This cutoff is sufficiently small to contain stars in the thick disk component whose circularity values are typically centered around $\epsilon\sim0.5$--$0.6$ \citep{Abadi2003SimulationsDisks,Obreja2018NIHAOHaloes}.
Some other studies have instead applied more strict cutoffs, e.g., $\epsilon=0.7$ or $0.8$, \citep[e.g,][]{Marinacci2014TheSimulations,Rodriguez-Gomez2016TheStars} for extracting purely rotating thin disk components with less contamination from the spheroidal component.
Our qualitative conclusion, however, does not depend critically on the choice of the cutoff.
Since our main focus is the early history of galaxy evolution (down to $z=0.7$) where the distinct two structures begin to appear in a galaxy, we use a simple cutoff instead of adopting more sophisticated decomposition techniques \citep[e.g.,][]{Obreja2018IntroducingGalaxy}.
Thus, we define the ``spheroidal components'' as all \replace{the stellar particles with disordered orbits ($\epsilon <0.5$).}

\begin{figure}[t]
    \linespread{1.0}\selectfont{}
    \includegraphics[width=\columnwidth]{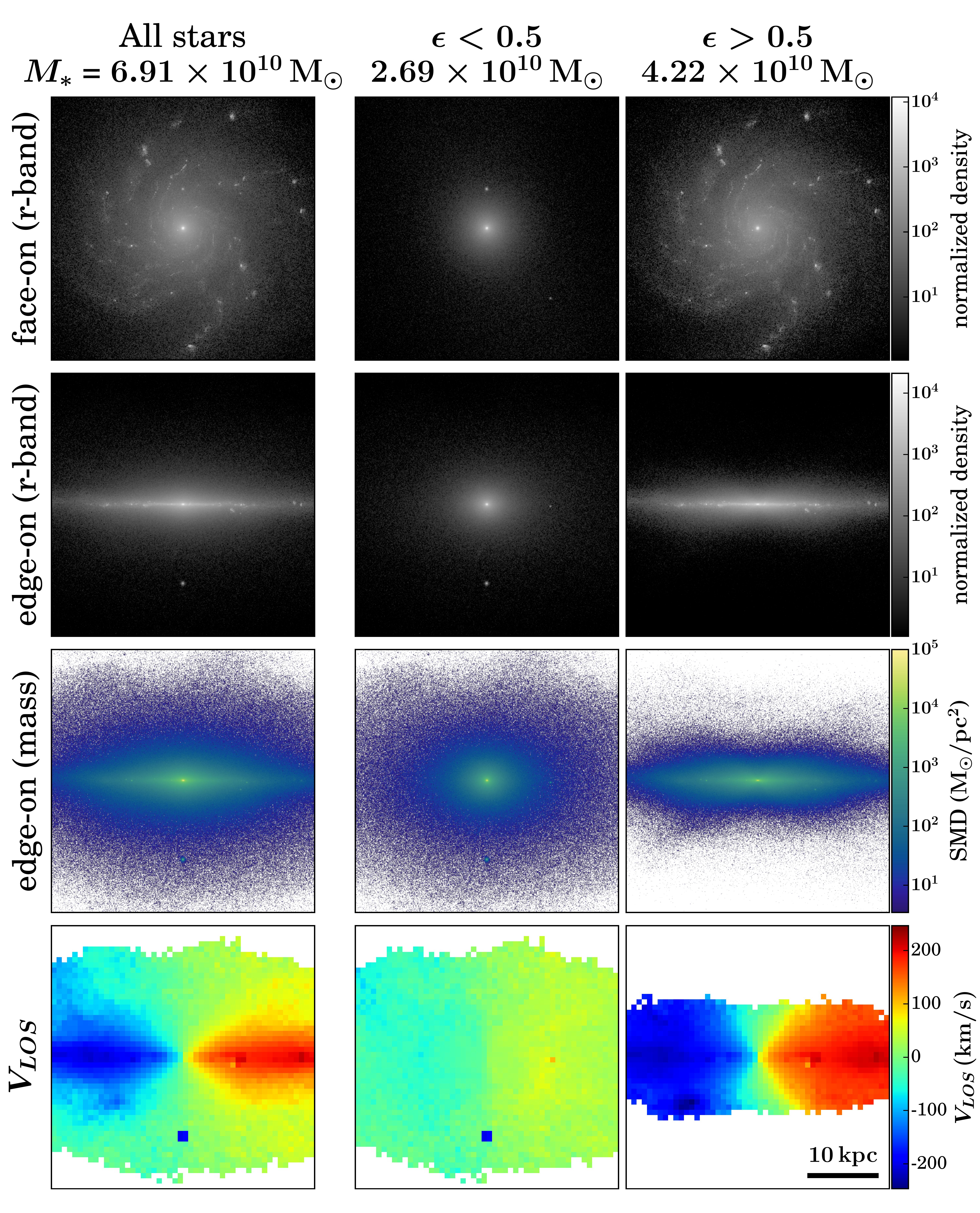}
    \caption{The images of a sample disk-dominated galaxy (left) and its decomposed structures: the disk (middle) and spheroidal (right) components. 
    The rest frame $\sl r$-band images in face-on (top row) and edge-on (second row) views. 
    The mass distributions in an edge-on view (third row), and the $\sl r$-band flux weighted line-of-sight velocity map in an edge-on view (fourth row). 
    We confirm a distinction between the kinematically decomposed structures, disk and spheroid, in the visual appearance and velocity maps. The radius of each box is $R_{90}$ of the galaxy, and the velocity map is displayed on the 50$\times$50 grid. Only the pixels containing more than 100 stellar particles are shown on the map.} 
    \label{fig:decomposed}
\end{figure}

Figure~\ref{fig:decomposed} shows the images of one of the most massive disk-dominated galaxies and its two components decomposed based on the circularity parameter: disk ($\epsilon \geq 0.5$) and spheroid ($\epsilon<0.5$). 
The top two rows are $\sl r$-band--weighted images (in the rest frame, without dust extinction) of the galaxy and its structures viewed face-on and edge-on respectively, while the third row shows the mass--weighted images. 
The last row is the line-of-sight velocity map in the  edge-on direction, weighted by the $\sl r$-band flux of the stars in each bin. 
The $\sl r$-band is calculated for every stellar particle based on stellar age and metallicity, following the stellar population synthesis models of \cite{Bruzual2003Stellar2003}. 
Stars with $\epsilon<0.5$ are indeed distributed smoothly as expected for the spheroidal component, whereas the disk stars with $\epsilon \geq 0.5$ show luminous (young) spiral-arm structures in the face-on view while looking thin in the edge-on view. 
In the line-of-sight velocity map, the difference between the components is also evident; 
stars with $\epsilon<0.5$ build a dispersion-dominated system with little net rotation because their line-of-sight velocities are canceled out, while disk stars show a clear rotation with $V_{rot}\sim200\,\rm km/s$.

\subsection{Kinematic morphology ($D/T$) \\- morphological mix in the field environment}
\label{sec:3.2}

\begin{figure*}
    \linespread{1.0}\selectfont{}
    \includegraphics[width=\textwidth]{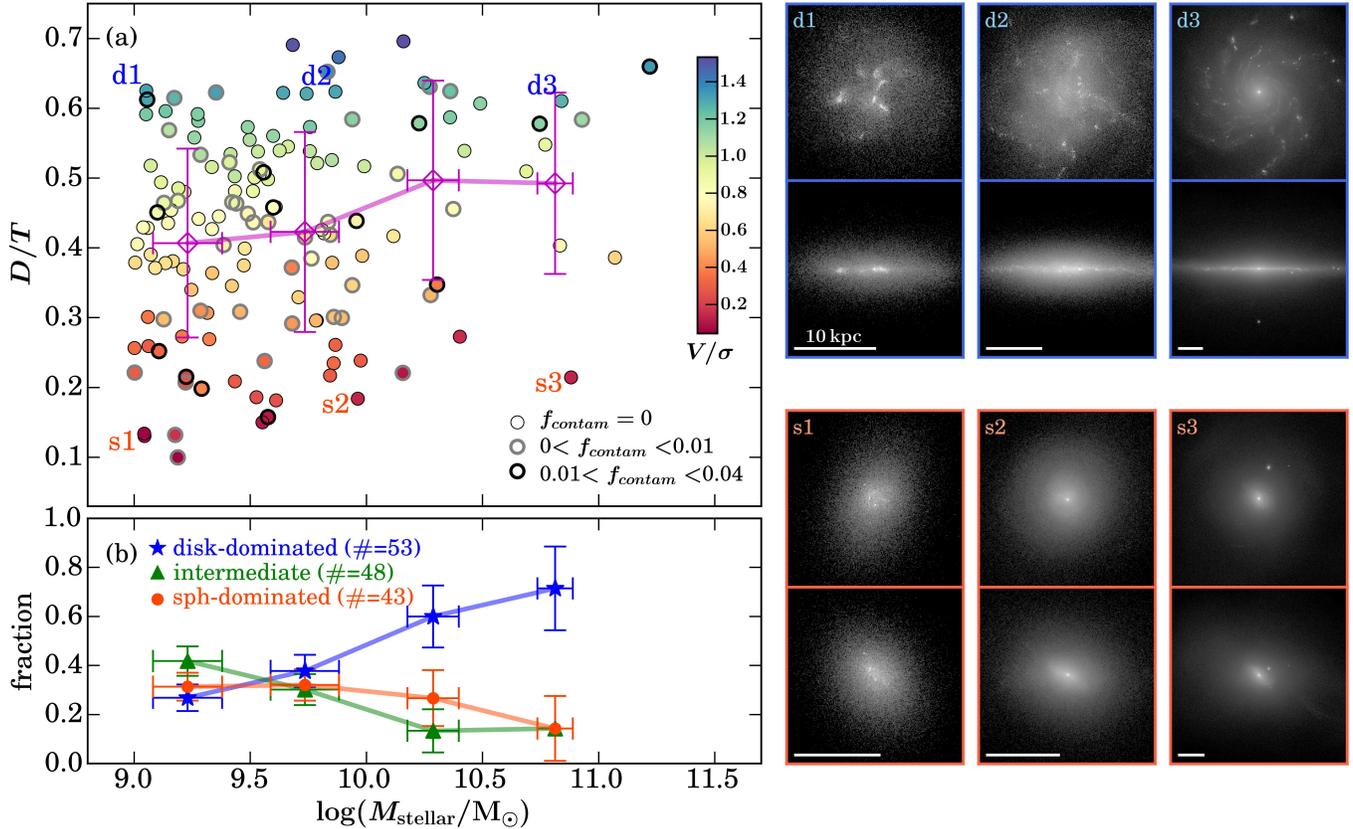}
    \caption{(a) $D/T$ as a function of galactic stellar mass (at $z=0.7$). 
    The colors represent the $V/\sigma$ of the galaxies. The \replace{magenta} diamonds and error bars represent the average $D/T$ and the standard deviation, respectively. 
    (b) The fraction of disk-dominated ($D/T>0.5$, blue stars), intermediate ($0.35<D/T<0.5$, green triangles), and spheroid-dominated ($D/T<0.35$, red circles) galaxies as a function of galactic stellar mass. 
    The error bars are measured from the standard error of the mean of a binomial distribution.
    The face-on and edge-on $\sl r$-band images of the three of the disk-dominated (d1, d2, d3) and spheroid-dominated (s1, s2, s3) galaxies are shown as examples on the right panels.
    We find that the average $D/T$ and the fraction of the galaxies with different morphology depend on the stellar mass.} 
    \label{fig:m_dt}
\end{figure*}

The morphology of galaxies can be approximated by the mass ratio of the disk components to the total stellar mass, $D/T$.
Figure~\ref{fig:m_dt}(a) shows the disk-to-total ratios of the New Horizon galaxies at $z=0.7$ as a function of the stellar mass.
The average values of $D/T$, \replace{represented as the magenta diamonds}, seem to show a slight positive trend with the stellar mass of the galaxies.
As a check, we also measured $V/\sigma$ of the galaxies, the degree of rotational support.
First we define the cylindrical coordinate system with respect to galactic spin axis (z-axis) and compute the radial ($V_{r}$), tangential ($V_{t}$), and vertical ($V_{z}$) component of velocity. 
Velocity dispersion is measured with respect to the mean value of each component (i.e. $\sigma_{r,t,z}^2= \bracket{V_{r,t,z}^2} - \bracket{V_{r,t,z}}^2$).
Using all stellar particles in a galaxy, $V/\sigma$ is calculated by the mean rotational velocity ($V=\bracket{V_{t}}$) normalized by the 1D-mimicking velocity dispersion:
   $\sigma = \sqrt{(\sigma_{r}^2+\sigma_{t}^2+\sigma_{z}^2)/3}$.
As suggested in the gradual color variation of $D/T$ (representing $V/\sigma$), we confirm that our $D/T$ measurement correlates well with $V/\sigma$ \footnote{\replace{Note that $V/\sigma$ here is different from observed $V/\sigma$; while we consider all the stellar particles in a galaxy in the measurement of $V/\sigma$, many of the observational studies measure $V/\sigma$ from the velocity moments maps within the projected half light radius \citep[e.g.,][]{Emsellem2007TheGalaxies}. We note that $V/\sigma$ presented here is not for a direct comparison with observations but for a sanity check on the reliability of $D/T$ measurements and confirming the validity of the trends.}}.

To examine the differences in the kinematically traced morphology more carefully, we classified the galaxies with $D/T$ larger than 0.5 as disk-dominated galaxies and those with $D/T$ smaller than 0.35 as spheroid-dominated galaxies. 
The rest are called intermediate galaxies.
Disk-dominated galaxies make up 37\% of the sample (53/144), whereas 30\% of the sample (43/144) is spheroid-dominated. 
The face-on and edge-on $\sl r$-band images of the three of the disk-dominated (d1, d2, d3) and spheroid-dominated (s1, s2, s3) galaxies are shown as examples on the right panels in Figure~\ref{fig:m_dt}.
The fractions of disk-dominated (blue), intermediate (green), and spheroid-dominated (orange) galaxies in each mass bin are shown in panel (b).

It is interesting to note that, unlike the average $D/T$ (panel (a)), the fraction of disk-dominated galaxies rather dramatically increases with stellar mass.
The strength of the trend, however, is somewhat sensitive to the choice of $D/T$ cuts for disk and spheroid-dominated galaxies.
The difference in the populations between disk-dominated and spheroid-dominated galaxies begins to appear at $\sim10^{10}\,\msun$ and becomes more pronounced at higher masses. 
This morphology-mass trend is consistent with the observational studies of \citet{Kassin2012TheNow} (see also \citealp{Simons2017Z2:Assembly,Johnson2018TheGalaxies}), while it may seem contradictory to what we observe in the local Universe, where more massive galaxies tend to be earlier type \citep{Conselice2006TheSystem}. 
This discrepency can be explained by noting that our sample is limited to the field environment, and the simulation has not reached $z=0$.

\section{The origin of disk and spheroidal components} 
\label{section4} 

Understanding the formation of the kinematic components requires tracking their stellar origins - when and where stars come from. 
In this section, we sorted the origins of the stellar particles in each component depending on their birthplace and orbits at birth: stars born \insitu\, with ``aligned'' (co-rotating co-planar initial orbits) and ``misaligned'' orbits (counter-rotating or non co-planar initial orbits), and stars formed \exsitu\ and later accreted.
Based on this classification, we aim to address the questions raised in the introduction  regarding the formation of disk and spheroidal components.

\subsection{The origins of stellar particles:\\ \insitu/\exsitu\, and birth orbital properties}
\label{sec: 4.1}
\begin{figure*}
    \linespread{1.0}\selectfont{}
    \includegraphics[width=\textwidth]{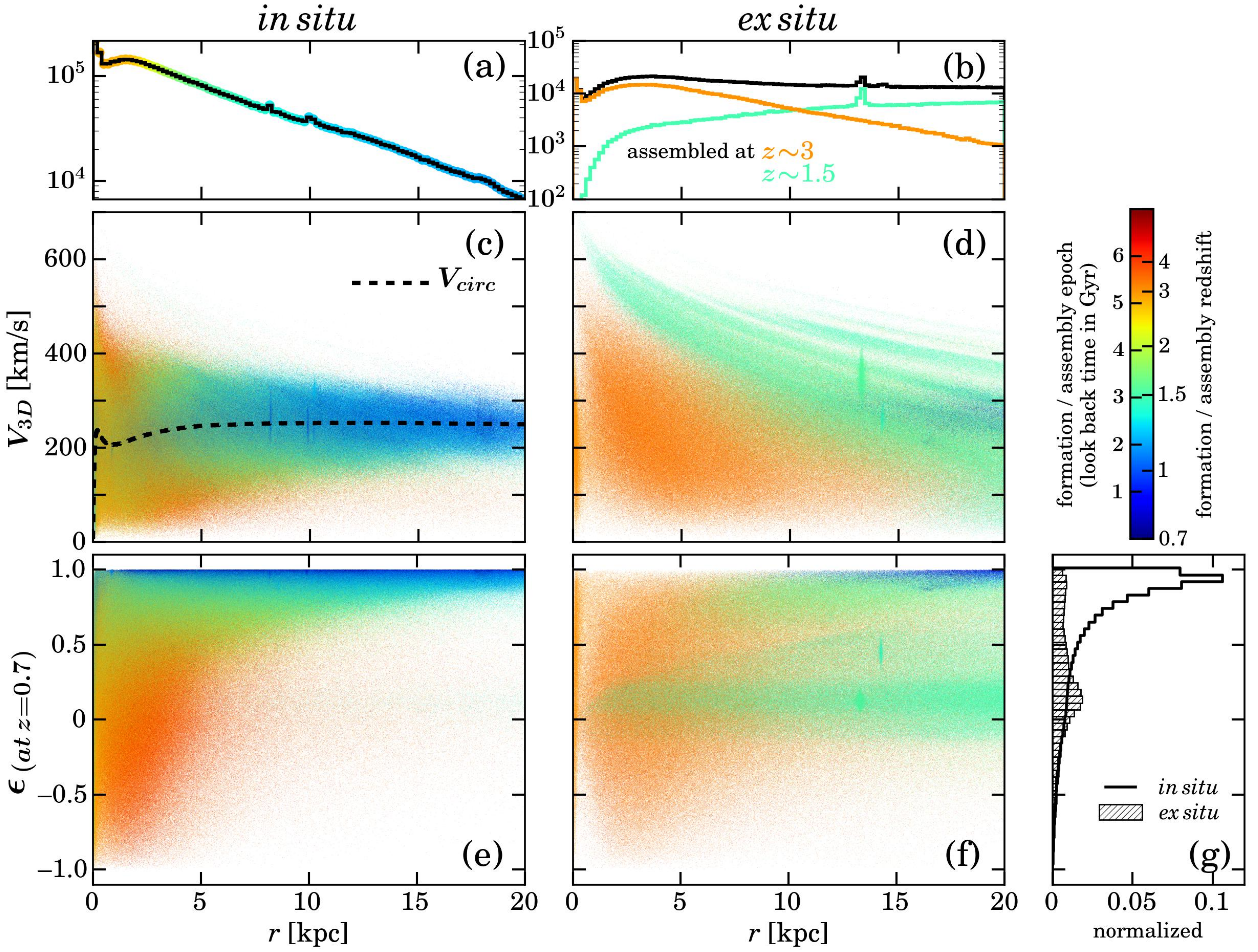}
    \caption{Distribution of the \insitu\ and \exsitu\ stars in a galaxy in a phase-space diagram. 
    \replace{(Top panels) The number of \insitu\, and \exsitu\, stars as a function of galactocentric distance. The distribution of \insitu\, stars are colored by median formation epoch of the stars at given distances. Orange and cyan lines are distributions of \exsitu\, stars assembled at $z\sim3$ and $z\sim1.5$ (two significant accretions through mergers), respectively.}
    (\replace{Middle} panels) 3D velocity versus the distance from the galactic center. 
    (Bottom panels) The circularity parameter versus the distance. 
    Each star is color-coded according to its \replace{formation (\insitu) or assembly (\exsitu) epoch, from red (early) to blue (recent)}. 
    The rotational velocity curve of the galaxy is plotted as the dashed line in panel (c). 
    (g) PDFs of the circularity parameters of \insitu\ (unfilled) and accreted (hatched) stars normalized by the total number of stars in the galaxy.
    The stellar kinematics is strongly dependent on the \replace{formation/}assembly epochs and origins; young stars formed \insitu\ have ordered motions with circularity close to 1, while stars formed earlier have disordered motions with low circularity. 
    Accreted stars have mostly disordered motions with circularity centered around 0, and their radial distributions depend on their assembly epoch.} 
    \label{fig:psd}
\end{figure*}

The origin of stars can be quantified as a function of where they formed: \insitu/\exsitu.
We tracked all the progenitors of the galaxy (up to at least $z=4$).
For each galaxy of the $z=0.7$ sample, stars born in its main progenitors are considered \insitu\ stars, and the rest as \exsitu\ stars\footnote{In each snapshot from (at least) $z=4$ to $0.7$, stars younger than $\rm 50\,Myr$ located within $R_{90}$ of its main progenitor are labelled as \insitu\ stars, and those stars that have not been labelled at all (meaning that they are born in the other progenitors and accreted later when they are older than $\rm 50\,Myr$) are considered to be \exsitu\ stars.}.
\replace{Figure~\ref{fig:psd} shows the distribution of stars in the same disk-dominated galaxy of Figure~\ref{fig:decomposed}.}
\replace{The top two rows show the number of \insitu\,(panel (a)) and \exsitu\ (panel (b)) stars as a function of distance. The distribution function of \insitu\, stars are colored according to the median formation epoch of the stars at given distances. The distribution of \exsitu\, stars assembled at $z\sim3$ and $z\sim1.5$ ($\rm 0.5\,Gyr$ window) are also added in panel (b).}
\replace{In the middle panels, stars are distributed} in the plane of the 3D velocity and the galactocentric distance, while the bottom rows show the circularity parameter versus distance.
\replace{Stars in each panel are color-coded according to their formation (\insitu) or assembly epoch (\exsitu).}

It is clear from Figure~\ref{fig:psd} that the present stellar kinematics is strongly dependent on the \replace{formation/}assembly epoch and origin. 
The gradual color variations in panels \replace{(c)} and \replace{(e)} indicate that the stars formed \insitu\ have different kinematics depending on their ages. 
The stars formed recently have circularity values close to 1, following the circular rotation curve, which is represented as the dashed line in panel \replace{(c)}. 
Therefore, the majority of young stars in the galaxy are formed in the disk.
Conversely, old stars formed \insitu\ are mostly contained in the inner regions with lower circularity, many of which comprise the bulge component.

In the phase-space diagram for \exsitu\ stars, \replace{Figure~\ref{fig:psd}(d) and \ref{fig:psd}(f)}, we can identify different families of accreted stars coming from different merger events. 
This particular galaxy undergoes significant accretions at $z\sim3$ \replace{(orange)} and $z\sim1.5$ \replace{(cyan)} through mergers, which can be seen as different groups of accreted stars with different assembly epochs.
Unlike the stars formed \insitu\ whose circularity parameters are skewed towards $\epsilon=1$, accreted stars mostly have disordered motions with circularity parameters almost centered around $\epsilon\sim0$ (Figure~\ref{fig:psd}(g)). 
Moreover, the distribution of accreted stars in the phase-space plane varies with epochs of assembly, as has been addressed by \cite{Font2011CosmologicalGalaxies}.
The stars from a recent accretion (cyan stars in Figure~\ref{fig:psd}(d) and \ref{fig:psd}(f), assembled at $z\sim1.5$) are distributed at large distances from the center with higher velocities, while stars accreted earlier (orange stars, assembled at $z\sim3$) have settled down toward the central part of the galaxy, because later accretion carries higher angular momentum (e.g.,~\citealp{White1984AngularProtogalaxies,Kimm2011TheRevisited,Stewart2011OrbitingAccretion}).

The most visible feature in this diagram is perhaps that stars that formed earlier tend to contribute to the spheroid and to have more disordered orbits.
The orbital characteristics of a star, however, are not fixed over time. 
The ordered orbits of disk stars can be disturbed as stars exchange angular momentum through mergers or instabilities.
Therefore, stars in each component do not remain where they are born.
To determine which component (disk/spheroid) stars belonged to when they were born, and therefore to quantify the amount of migration from disks to spheroids, we measured the orbital properties of the \insitu\ stars at birth.
The stars born \insitu\ were divided again based on their circularity parameter at birth.
The same choice of circularity cutoff $\epsilon_{birth}=0.5$ was employed to determine whether a star was born a disk star.
In other words, if the initial orbit of a star (at birth) is ``aligned'' with the co-rotating disk plane ($\epsilon_{birth}>0.5$), then it means that it was born a disk star. 
Otherwise ($\epsilon_{birth}<0.5$, ``misaligned orbits'' with the co-rotating disk plane), it is most likely to have formed from misaligned, unsettled gas, likely during a merger.

\begin{figure*}[t]
    \linespread{1.0}\selectfont{}
    \includegraphics[width=\textwidth]{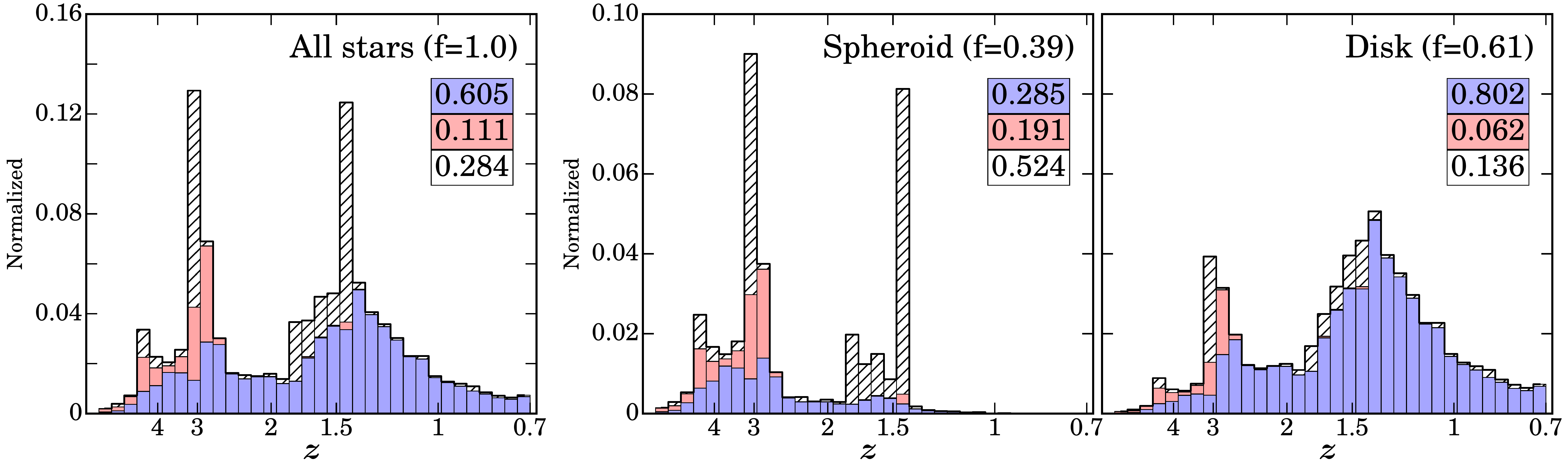}
    \caption{Assembly history of the stars in a galaxy selected at $z=0.7$ (with mass of $\rm 6.91\times10^{10}\,\msun$, \replace{the same galaxy of Figure~\ref{fig:decomposed}}) and its two components.
    Normalized histograms of the assembly epochs of all the stars in the galaxy (left), stars in the spheroidal (middle) and disk (right) components.
    The stacked hatched bars represent the number of stars accreted at each epoch. 
    The stars born \insitu, stacked underneath, are divided into two groups depending on their orbits at birth: aligned (blue) and misaligned (red) orbits. 
    The fractions of each subcomponent in the spheroid and disk are shown in the upper right box.} 
    \label{fig:assembly}
\end{figure*}

The leftmost panel in Figure~\ref{fig:assembly} shows the assembly history of \replace{the same disk-dominated galaxy of Figure~\ref{fig:decomposed}} with stellar mass of $\rm 6.91\times10^{10}\,\msun$.
Each bin represents the number of stellar particles born \insitu\ with aligned (blue) and misaligned (\replace{red, on top of blue}) orbits and the number of stellar particles formed \exsitu\, and later accreted by this galaxy (\replace{hatched, on top of red}).
The boxes in the top right corner of the panel show the fractions of these subcomponents in the galaxy. 
As indicated by colors in Figure~\ref{fig:psd}, this galaxy experienced significant accretion events at $z\sim3$ and $z\sim1.5$ through mergers, resulting in \replace{the two maxima of stars accreted at these redshifts (spikes in the hatched histograms). The total \exsitu\ mass fraction is $\sim0.284$.}
At high redshifts, most stars form from misaligned and unsettled gas (red bars) and, later, disk stars (blue bars) start to form predominantly. 
The general trend of the star formation responsible for disk stars will be discussed in Section~\ref{sec: 4.2}.

The two panels on the right, in the same format as the left panel, show when and how the stars constituting the spheroid (middle) and the disk (right) at $z=0.7$ are assembled into the galaxy.
The spheroidal component consists of accreted stars \replace{(accounting for 52.4\% of the spheroid)}, stars born with misaligned orbits \replace{(19.1\%)}, and stars initially born in the disk (\replace{initial} aligned orbits) yet migrated to the spheroid as their orbits are perturbed \replace{(28.5\%)}.
The disk component, conversely, is primarily composed of the stars formed \insitu\ with aligned orbits (more than 80\%, for this galaxy). 
The distribution of the assembly epochs of the stars in each component implies that the stars in the spheroidal component are relatively older than the disk stars because the galaxy continuously produces young disk stars. 
We further calculated the median age of the spheroidal and disk component of this galaxy at $z=0.7$ and confirmed that the spheroid stars ($\rm 5.35\,Gyr$) are much older than the disk stars ($\rm 3.01\,Gyr$).
\replace{The qualitative assembly history of disk and spheroid and their origins (high \exsitu\, fraction for spheroidal components, mainly for the stellar haloes given their distribution in Figure~\ref{fig:psd}) agree with the results of \cite{Obreja2018NIHAOHaloes}.}

\subsection{Formation of disks}
\label{sec: 4.2}
As implied in Figure~\ref{fig:assembly}, disk stars tend to form at later epochs, while early star formation contributes to the growth of the spheroidal components.
In this section, we explore when the galaxies start to predominantly form disk stars. 
Also, we measure how many of the stars in a galaxy are born as disk stars and investigate how this correlates with galactic stellar mass, final morphology, and accreted fraction.

\subsubsection{When do galaxies start to predominantly form \\ disk stars? -- disk-mode star formation}
\label{sec: 4.2.1}

\begin{figure*}
    \linespread{1.0}\selectfont{}
    \includegraphics[width=\textwidth]{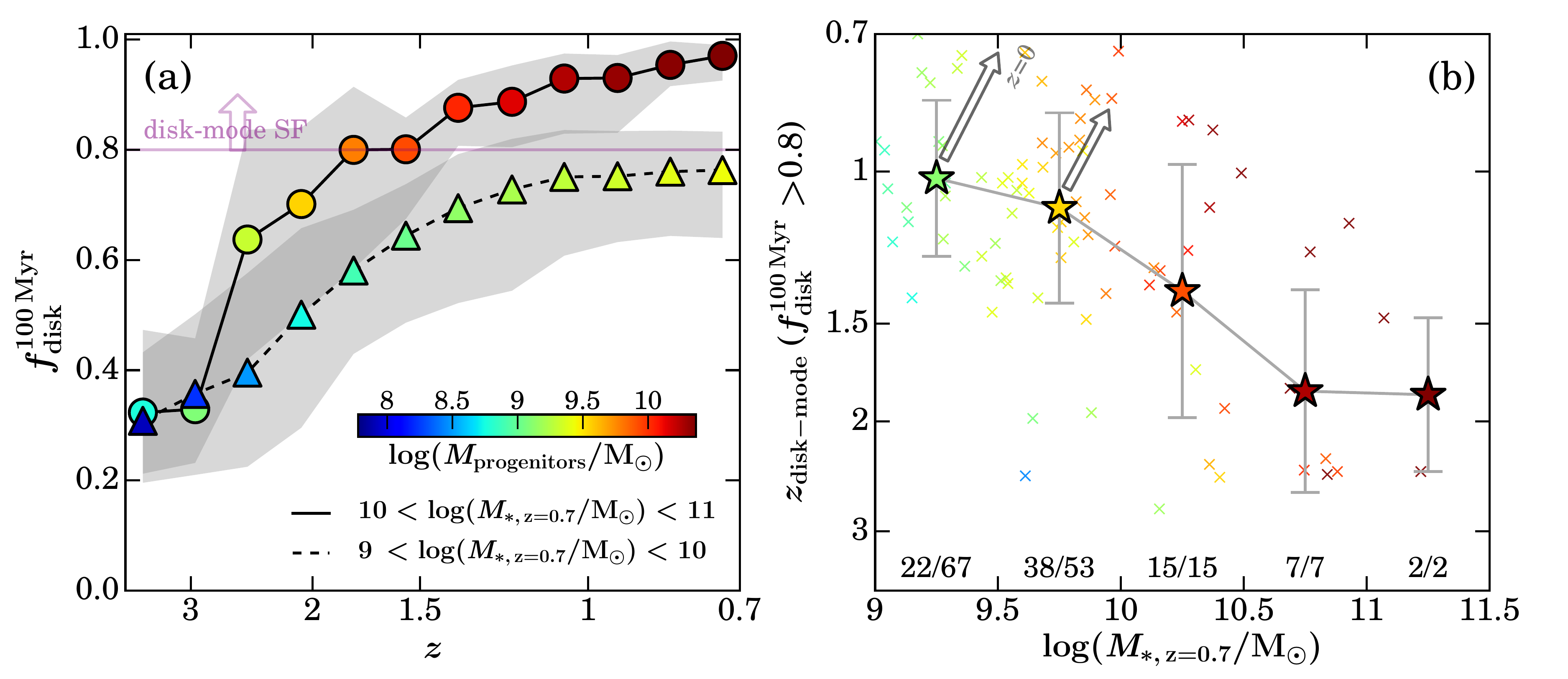}
    \caption{(a) The fraction of disk stars among the stars formed over a time period of $\rm 100\,Myr$ (\fyoungdisk) as a function of redshift. 
    The solid line with circles shows the evolution of median \fyoungdisk\, for massive galaxies ($\rm 10<log(\it M_{*,\rm z=0.7}/\rm M_{\odot})<11$), and the dashed line with triangles shows the evolution of low-mass galaxies ($\rm 9<log(\it M_{*,\rm z=0.7}/\rm M_{\odot})<10$).
    The two groups of galaxies are divided according to their stellar mass at $z=0.7$, and the colors inside the markers represent the median stellar mass of their main progenitors at the corresponding redshift.
    The gray shadings represent the 30th and 70th percentiles. 
    (b) The $z_{\rm disk-mode}$, the epoch at which galaxies begin to form stars in  ``disk-mode'' (i.e., $f_{\rm disk}^{\rm 100\,Myr}>0.8$ for the last $\rm 300\,Myr$), as a function of stellar mass of galaxies.
    Each star represents the epoch averaged over the galaxies in the corresponding mass bin, and \replace{the error bars show the standard deviations}. 
    The colors inside the stars are, again, coded by the average stellar mass of the progenitors at $z_{\rm disk-mode}$ (i.e., $M_{\rm disk-mode}$).
    The bottom numbers indicate the number of galaxies that have ever undergone the disk-mode star formation by $z=0.7$ among the total number of galaxies in the mass bin. 
    Thus, only a small fraction of the low-mass galaxies are taken into account in the measurement of $z_{\rm disk-mode}$. If we assume that those galaxies form stars in disk-mode after $z\sim0.7$ so that we count all of the low-mass galaxies in the calculation, the estimated $z_{\rm disk-mode}$ for the two lowest mass bin will likely to move up-and-rightward as marked by arrows (scale and direction are arbitrary). 
    We find that more massive galaxies (e.g., $M_{\rm*,z=0.7}>10^{10}\,\msun$) start to form disk stars from earlier epochs ($z\sim1-2$).}
    \label{fig:diskmode_sf}
\end{figure*}

In order to quantify the kinematic properties of newly formed young stars at each redshift, we measure the fraction of disk stars among the stars formed over a time period of $\rm 100\,Myr$, \fyoungdisk.
Figure~\ref{fig:diskmode_sf}(a) shows \fyoungdisk\, as a function of redshift for two groups of galaxies divided based on their stellar mass at $z=0.7$; the solid line with circles shows the evolution of the median \fyoungdisk\, for massive galaxies ($\rm 10<log(\it M_{\rm *,z=0.7}/\rm M_{\odot})<11$), and the dashed line with triangles shows the evolution of low-mass galaxies ($\rm 9<log(\it M_{\rm *,z=0.7}/\rm M_{\odot})<10$).
Each point is the median value of \fyoungdisk\, measured at the corresponding redshift on x-axis with a time window of $\rm{\pm150\,Myr}$.
The colors inside the markers represent the median stellar mass of their main progenitors at the corresponding redshift.

As shown in Figure~\ref{fig:diskmode_sf}(a), both lines increase with time. 
This trend is expected for two reasons; as redshift decreases, the mergers become less frequent \citep[e.g.,][]{Rodriguez-Gomez2015TheModels, Rodriguez-Gomez2016TheStars}, and the angular momentum of accreted material increases \citep[e.g.,][]{Kimm2011TheRevisited}. 
The most notable feature in this trend is that \fyoungdisk\, increases more steeply in massive galaxies (the solid line with circles), meaning that more massive galaxies start to form disk stars at earlier epochs. 
This trend qualitatively agrees with the observations of  \citet{Kassin2012TheNow} \citep[see also][]{Simons2016KinematicZ2,Simons2017Z2:Assembly, Johnson2018TheGalaxies}.
Based on gas kinematics, they measured the fraction of galaxies with settled disks (probed by $V/\sigma$) and found that the fraction increases with stellar mass and time.
Recent numerical studies have also reproduced this trend. 
For example, \cite{El-Badry2018GasSimulations} used 24 FIRE galaxies with stellar masses of $\rm 6<log(\it M_{*}/\rm M_{\odot})<11$ and found a mass dependence on the formation of rotationally-supported gas disks. 
In addition, \cite{Pillepich2019FirstTime} have shown a similar trend that more massive galaxies have thinner disks, in the sense of better defined morphological structure and kinematics, than lower-mass galaxies.
This disk settling issue will be discussed in more detail in a separate paper (Dubois et al. in prep.).

Another noticeable feature of Figure~\ref{fig:diskmode_sf}(a) is that high-redshift galaxies with settled disks (progenitors of the massive galaxies \replace{at $z=0.7$}) are heavier than the galaxies whose disks settled more recently; the colors inside the markers, representing the median progenitor mass at the corresponding redshift, are redder to the left at fixed  \fyoungdisk\, (in the horizontal direction).
In order to investigate the mass associated with disk settling more carefully, we define the disk-mode star formation as when \fyoungdisk\, is higher than $0.8$\footnote{The resulting qualitative trend does not depend sensitively on the choice of cutoff for the disk-mode star formation. We, however, take the high cutoff of \fyoungdisk=0.8 to prevent any contamination that might be caused by our simple way of distinguishing disk-born stars ($\epsilon_{birth}>0.5$).} for the last $\rm 300\,Myr$. 
We measure \qsay{when} galaxies \qsay{begin} this disk-mode star formation (disk-mode epoch, $z_{\rm disk-mode}$).
Figure~\ref{fig:diskmode_sf}(b) shows $z_{\rm disk-mode}$ as a function of stellar mass of galaxies at $z=0.7$.
Each star represents $z_{\rm disk-mode}$ averaged over the galaxies in the corresponding mass bin, and the color is coded by $\it M_{\rm disk-mode}$, the median mass of the progenitors at $z_{\rm disk-mode}$.

The trend in Figure~\ref{fig:diskmode_sf}(b) clearly shows an anti-correlation between galactic stellar mass and disk-mode epoch; massive galaxies ($\rm 10<log(\it M_{*,\rm z=0.7}/\rm M_{\odot})<11$) start to form disks from $z\sim1-2$ \citep[e.g.,][]{Simons2016KinematicZ2, Hung2019WhatGalaxies}, \replace{when their mass become $\sim10^{10}\,\msun$ \citep[e.g.,][]{Clauwens2018TheFormation,Trayford2019TheSimulations}}, while low-mass galaxies ($\rm 9<log(\it M_{*,\rm z=0.7}/\rm M_{\odot})<10$) do after $z\sim1$.
In addition, $M_{\rm disk-mode}$ (the colors inside the markers) seems to mildly decrease with decreasing $z_{\rm disk-mode}$, which \replace{can be interpreted as a result of decreasing violence of the environments at lower redshifts. At higher redshifts, where frequent star forming activity, mergers, and gas accretion trigger gas turbulence \citep[e.g.,][]{Genel2012OnDiscs,Turner2017The3.5}, only massive galaxies in deep gravitational potential wells are able to form a majority of stars in ``disk-mode''. As redshift decreases, the environments become less violent, and galaxies with lower mass start the disk-mode star formation.}


One thing we need to also consider is that not all low-mass galaxies are taken into account in the measurement of this disk-mode epoch.
The numbers at the bottom of panel (b) of Figure~\ref{fig:diskmode_sf} indicate the number of galaxies that have ever undergone the disk-mode star formation by $z=0.7$ out of the whole galaxy sample in the mass bin.
For instance, only around a third (22/67) of the galaxies with masses between $10^9$ and $10^{9.5}\,\msun$ appear in this figure (each is marked as \qsay{$\times$}), and the rest of them have never been forming stars in ``disk mode''.
Therefore, assuming that those galaxies will start the disk-mode star formation at some point between  $z=0.7$ and $z=0$, eventually in the local Universe, the two points in the lowest mass bins will be likely higher if measured at $z=0.0$, while the other points may not move much. 
The galaxies in the two lowest mass bins will probably grow in mass as well; and thus the two lowest mass data points in this diagram will likely to move up-and-rightward as marked by arrows, while the exact direction and magnitude are unclear at the moment.
As a result, we expect a steeper trend between the stellar mass and $z_{\rm disk-mode}$ at $z=0$ than what appears here at $z=0.7$.

\subsubsection{How many stars are born as disk stars in total?}
\label{sec:4.2.2}
\begin{figure}[h]
    \linespread{1.0}\selectfont{}
    \centerline{\includegraphics[width=\columnwidth]{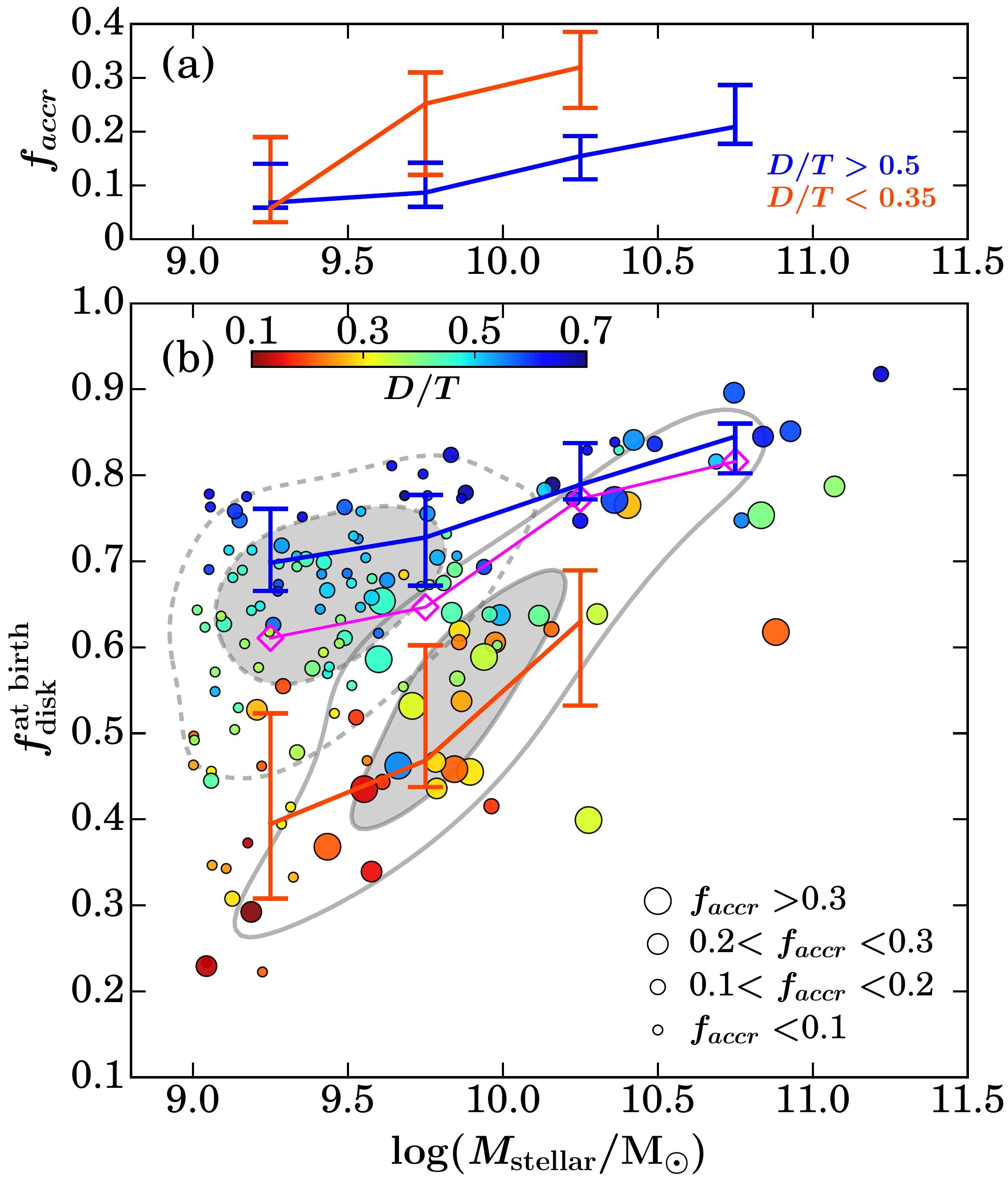}}
    \caption{(a) Median fraction of accreted (\exsitu) stars in the disk-dominated (blue line) and spheroid-dominated (orange line) galaxies in each stellar mass bin.
    (b) The fraction of disk stars at birth (\fbirthdisk) as a function of galactic stellar mass. 
    \replace{The median \fbirthdisk\, in each mass bin is shown as the magenta line with diamonds.}
    Each galaxy is colored according to $D/T$ at $z=0.7$, and the size of the circle indicates the fraction of accreted stars.
    In other words, a larger circle implies that a galaxy had more significant mergers (in the sense of either number or impact) than a galaxy plotted as a small circle.  
    The solid and dashed contours show (shaded: $0.5\sigma$, unshaded: $1\sigma$) the distribution of galaxies with $f_{accr}$ higher and lower than 0.2, respectively, in the plane of $M_{\rm stellar}$--\fbirthdisk.
    The blue and orange lines represent the median \fbirthdisk\, for disk-dominated ($D/T>0.5$) and spheroid-dominated ($D/T<0.35$) galaxies in each mass bin.
    All the error bars in this figure represent the 20th and 80th percentiles.
    The \fbirthdisk\, is strongly dependent on the stellar mass and accretion, and it is essential in determining the final kinematic morphology.} 
    \label{fig:m_fbirthdisk}
\end{figure}

The kinematic morphology, i.e., mass ratio of (co-rotating) disk to total, in this study is determined using all the stellar particles in a galaxy. 
Therefore, even if a galaxy has recently started disk-mode star formation, it cannot be classified as a \qsay{kinematically} disk-dominated galaxy, because the newly formed stars cannot yet outnumber the preexisting stars, which have disordered motions. 
As the global star formation rate peaks near $z\sim2$ \citep[e.g.,][]{Hopkins2006OnHistory}, the overall amount of disk-born stars should be driven by the stars formed at these epochs.
In this context, we measure the fraction of disk stars at birth (\fbirthdisk) to quantify the independent effects of star formation in developing disks and spheroids, excluding the effects of migration between the components and merger accretions.
The total disk fraction at birth\footnote{Note that we only considered stars with co-rotating and co-planar orbits as ``disk'' component when measuring $D/T$ (mass ratio of stars with $\epsilon>0.5$) and \fbirthdisk\,($\epsilon_{birth}>0.5$).  Thus, due to our definition, stars formed in the counter-rotating disks, whose orbits are still co-planar but counter-rotating (e.g., $\epsilon_{birth}<-0.5$), have ``misaligned'' initial orbits, which does not contribute to increasing \fbirthdisk. We will discuss the development of counter-rotating disks from the gas infall misaligned with the existing co-rotating disk plane after mergers in more detail in Section~\ref{sec:4.4}.} is defined as the total number of stars born as (co-rotating) disk stars among the stars formed \insitu, which is shown as a function of the stellar mass of the galaxies in Figure~\ref{fig:m_fbirthdisk}(b):
\begin{equation}
\label{eq:fbirthdisk}
f_{\rm disk}^{\rm at\, birth}  = 
\dfrac{\text{\insitu\ born stars with }  \epsilon_{birth}>0.5}{\text{Total number of \insitu\ born stars}}.
\end{equation}
Each galaxy is color-coded according to its $D/T$ at $z=0.7$, and the size of the circle indicates the fraction of accreted (\exsitu) stars. 

\replace{In Figure~\ref{fig:m_fbirthdisk}(b)}, we found that the fraction of disk stars at birth increases with mass \replace{(see the magenta line)}. 
This is because, as shown in Figure~\ref{fig:diskmode_sf}, the heavier the galaxy, the earlier it starts disk-mode star formation.
Therefore, most stars in massive galaxies ($>10^{10}\,\msun$) are born as disk stars (\fbirthdisk$\sim$0.76, on average), which is in agreement with the results of the numerical simulation by \cite{Garrison-Kimmel2018TheSimulations}. 
They found that 60--90\% of the stars in MW-mass galaxies at $z=0$ are born as disk stars. 
This mass dependence of the initial fraction of disk stars at birth is blurred by migration between the components and accretion, resulting in the weaker mass dependence on $D/T$ shown in Figure~\ref{fig:m_dt}(a).

Another notable point is that \fbirthdisk\, is also related to  $f_{accr}$, as can be seen in the two contours in Figure~\ref{fig:m_fbirthdisk}(b).
Galaxies with $f_{accr}>0.2$ (solid contour) lie diagonally across the \fbirthdisk--$M_{\rm stellar}$ plane, which can be attributed to the following two reasons;
(i) more massive galaxies have higher accretion (e.g., \citealp{Oser2010TheFormation,Lee2017FormationMass}) and (ii) galaxies with higher accretion have lower \fbirthdisk\, at fixed mass (specifically, in the mass range of $10^9$--$10^{10}\msun$). 
In regard to (ii), it seems that mergers or fly-bys can boost gas turbulence in galaxies, making galaxies less likely to form stars in an orderly fashion \replace{(lowering \fbirthdisk)}. 
The formation of non-disk stars ($\epsilon_{birth}<0.5$) can be boosted by the central starburst \citep[e.g.,][]{Hernquist1989TidalGalaxies}.

Based on the relationship between $M_{\rm stellar}$, \fbirthdisk, and $f_{accr}$, we can understand the different morphology of galaxies as follows.
The blue and orange line in Figure~\ref{fig:m_fbirthdisk} represents the median $f_{accr}$ (panel (a)) and \fbirthdisk\, (panel (b)) as a function of stellar mass for disk-dominated ($D/T>0.5$) and spheroid-dominated ($D/T<0.35$) galaxies, respectively. 
Disk-dominated galaxies, on average, have higher \fbirthdisk\, and lower $f_{accr}$ than spheroid-dominated galaxies at fixed stellar mass. 
This means that they have started to form disk stars from earlier epochs (thus, high \fbirthdisk) and experienced fewer violent events that can destroy their disks (hence, high $D/T$ at $z=0.7$, the final epoch).

The majority of the stars in the spheroid-dominated galaxies ($D/T<0.35$, orange lines), on the other hand, are originally formed with misaligned orbits \replace{(low \fbirthdisk)}.
This appears to be more pronounced in the lower-mass galaxies ($<10^{10}\,\msun$) that have not developed disks until $z=0.7$ (see also Figure~\ref{fig:diskmode_sf}).
Massive spheroid-dominated galaxies ($>10^{10}\,\msun$), conversely, form higher fraction of disk stars (although still lower than disk-dominated galaxies with similar masses). 
Given their distinctly high $f_{accr}$, it can be assumed that mergers have made a significant contribution to the development of their spheroids.

\subsection{Growth of spheroids}
\label{sec:4.3}
\begin{figure*}
    \linespread{1.0}\selectfont{}
    \includegraphics[width=\textwidth]{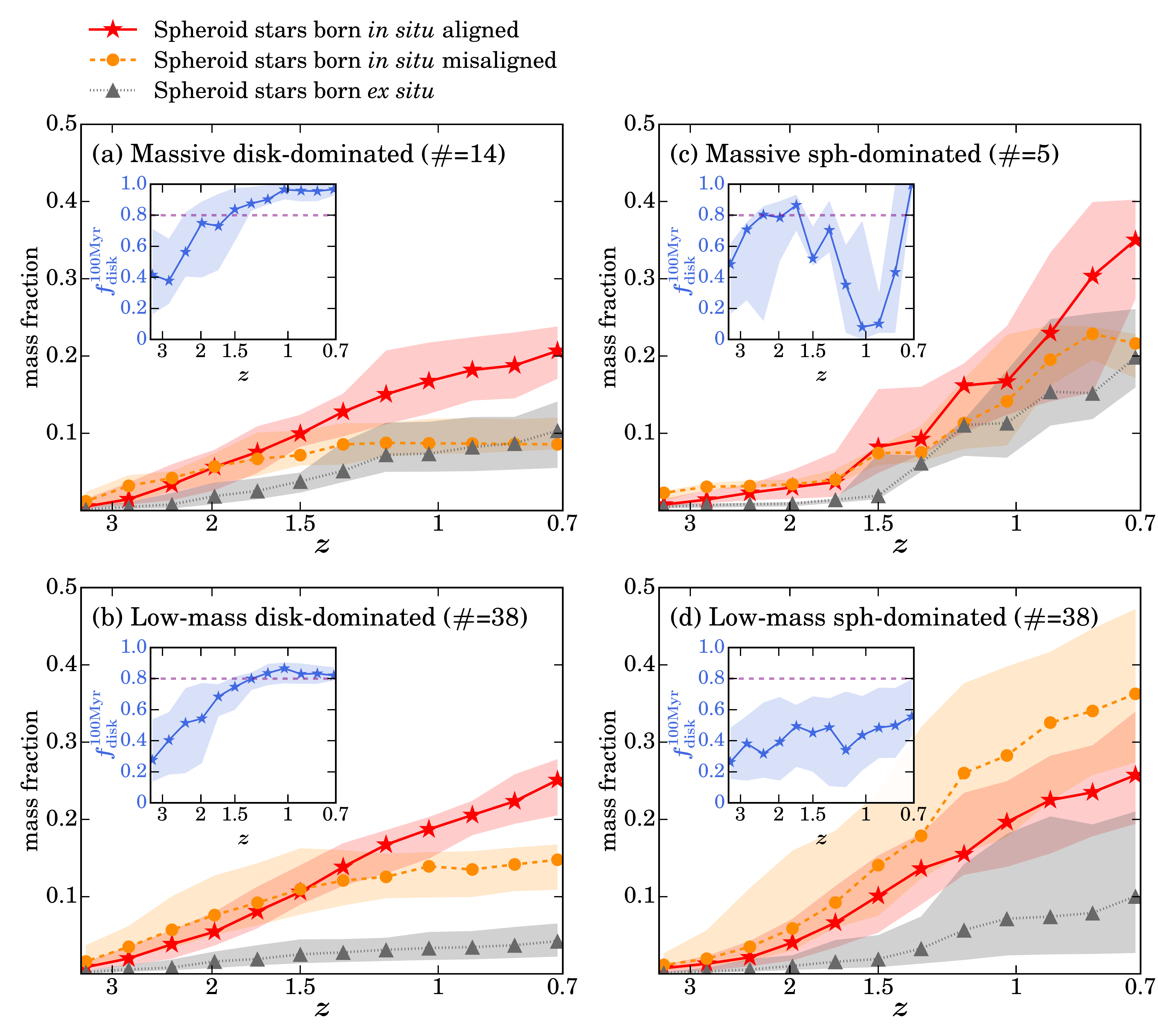}
    \caption{Evolution of the mass fraction of the subcomponents (measured with a time window of $\pm\rm50\,Myr$) in the spheroids to the final stellar mass at $z=0.7$.
    The three subcomponents are divided following Section~\ref{sec: 4.1}: stars born \insitu\ with aligned (\replace{red} line with stars) and misaligned (\replace{orange} line with circles) orbits and accreted stars (gray line with triangles).
    Galaxies are divided into four groups:
    (a) massive ($\rm 10<log(\it M_{\rm *,z=0.7}/\rm M_{\odot})<11$) disk-dominated galaxies, (b) low-mass ($\rm 9<log(\it M_{\rm *,z=0.7}/\rm M_{\odot})<10$) disk-dominated galaxies,
    (c) massive spheroid-dominated galaxies, and
    (d) low-mass spheroid-dominated galaxies.
    The number of galaxies in each group is shown in the parenthesis.
    Each inset panel shows the evolution of \fyoungdisk\ (defined as in Figure~\ref{fig:diskmode_sf}), and the horizontal dashed purple line represents the cut for disk-mode star formation \replace{(see Section~\ref{sec: 4.2})}. 
    All the shadings in this figure represent the 20th and 80th percentiles.
    A significant fraction of the spheroids comes from the disk stars that are perturbed (except for (d)), and the importance of accreted stars is greater in more massive galaxies (especially in (c)).}
    \label{fig:sph_growth}
\end{figure*}

Spheroids are believed to grow through several internal and external processes including early star formation, disk instability, and hierarchical merging. 
\replace{Indeed, \cite{Trayford2019TheSimulations} has pointed out that while stars formed at high redshifts generally contribute to the spheroidal components, transformational processes of disk stars formed at lower redshifts are also necessary to build the spheroid.}
Thus, quantifying the different origins of spheroid stars can provide hints about the importance of each formation process.
As described in Section~\ref{sec: 4.1}, we divided the origin of the stars according to their birthplace and birth orbital properties: \insitu\,-born disk stars (aligned birth orbits), \insitu\,-born spheroid stars (misaligned birth orbits), and accreted stars.
Figure~\ref{fig:sph_growth} shows the evolution of the mass fraction of each subcomponent in the spheroidal components to the total stellar mass at $z=0.7$. 
Galaxies are divided into four groups by \replace{their} final mass and morphology: (a) massive ($\rm 10<log(\it M_{\rm *,z=0.7}/\rm M_{\odot})<11$) disk-dominated galaxies, (b) low-mass ($\rm 9<log(\it M_{\rm *,z=0.7}/\rm M_{\odot})<10$) disk-dominated galaxies, (c) massive spheroid-dominated galaxies, and (d) low-mass spheroid-dominated galaxies.
Each inset panel shows the \replace{evolution of} \fyoungdisk\ (defined as in Section~\ref{sec: 4.2.1}), for the galaxies in each group.

In the disk-dominated galaxies (panels (a) and (b)), the growth of the spheroidal components is primarily driven by the increase of perturbed disk stars (\replace{red} solid lines).
As the inset panel shows, most of the stars formed at lower redshift are disk stars (\fyoungdisk$>0.8$, disk-mode star formation); therefore, their migration into the spheroidal components is the leading contributor to the growth of the spheroids, while the subcomponent consisting of the stars born with misaligned orbits (non-disk stars, \replace{orange} dashed lines) does not grow much after $z\sim1.5$.
This trend of increasing fraction of the spheroid stars that migrated from the disks agrees with the results from \cite{Zolotov2015CompactionNuggets}; their simulation which reached $z=1$ suggests that the fraction of these stars in spheroids is growing from $10\%$ at $z=5$ to $30\%$ at $z=1$ when averaged over 26 galaxies with masses higher than $10^{10}\,\msun$, although they used different circularity and spatial cut to extract the spheroidal (bulge) components.
As discussed in Section~\ref{sec:4.2.2}, accreted stars (gray dotted lines) seem to be more important in the more massive galaxies.

In the spheroid-dominated galaxies, each fraction of subcomponent at $z=0.7$ is almost two times higher than that of the disk-dominated galaxies with similar mass; for example, the spheroids with accreted stars (gray dotted line) account for $\sim20\%$ of the total stellar mass in the massive spheroid-dominated galaxies (panel (c)), whereas its contribution to the total stellar mass is only $\sim10\%$ in the massive disk-dominated galaxies (panel (a)).
As Figure~\ref{fig:m_fbirthdisk}(a) shows, spheroid-dominated galaxies have higher accreted fraction ($f_{accr}$) than disk-dominated galaxies at fixed stellar mass, and most of the accreted stars are more likely to build up the spheroidal components.

Especially, the spheroid-dominated galaxies has a larger contribution from non-disk stars (born with misaligned orbits, \replace{orange dashed} line) than disk-dominated galaxies.
Low-mass spheroid-dominated galaxies (panel (d)) have not formed stars in \qsay{disk-mode} at all (i.e., \fyoungdisk $<0.8$, throughout the history until $z\sim0.7$  as shown in the inset panel), and their spheroids grow mainly from the stars formed with misaligned orbits.  
On the other hand, massive spheroid-dominated galaxies (panel (c)) form stars in disk-mode already at $z\sim2$ (inset panel). 
The mass of their spheroidal components, however, dramatically increases after $z\sim1.5$, along with the sudden increase of accreted stars and the subsequent decrease of \fyoungdisk (see Section~\ref{sec:4.2.2}); the formation of stars with aligned orbits (disk stars) seems to be suppressed due to merger boosted gas turbulence.
Furthermore, disk stars dramatically migrate to the spheroidal components (\replace{red solid} line), which is very likely linked to the morphological transformation induced by mergers.

\subsection{Misaligned gas stream \\-counter-rotating structures} 
\label{sec:4.4}
\begin{figure}[h]
    \linespread{1.0}\selectfont{}
    \centerline{\includegraphics[scale=0.4]{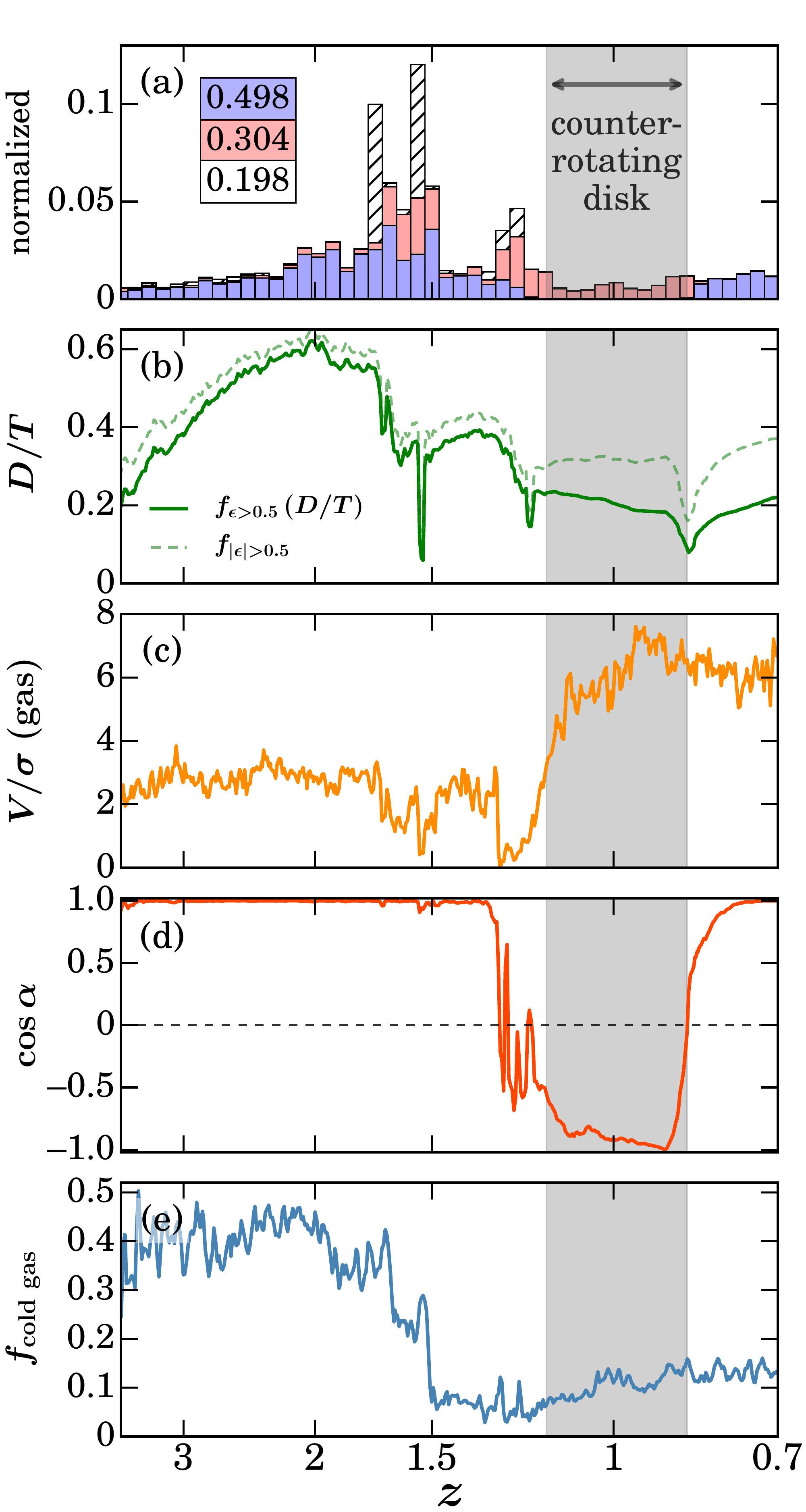}}
    \caption{Evolution of a galaxy developing a counter-rotating disk.
    (a) The normalized stacked histogram of the assembly epochs of all stars in the galaxy, color-coded according to the different origins of the stars in the same way as in Figure~\ref{fig:assembly}
    (blue: stars formed \insitu\ with $\epsilon_{birth}>0.5$;
    red: stars formed \insitu\ with $\epsilon_{birth}<0.5$; and hatched: stars formed \exsitu).
    The fraction of each subcomponent is shown in the upper right box. 
    (b) The evolution of $D/T$ (co-rotating disk fraction, $f_{\epsilon>0.5}$) as a function of the redshift. The dashed line shows the mass ratio of the stars with $|\epsilon|>0.5$ (co-planar orbits including both co- and counter-rotation).
    (c) The $V/\sigma$ of the cold gas inside the galaxy, measured with respect to the spin axis of the gas.
    (d) The cosine angle ($\rm cos\alpha$) between the rotational axes of the gas and stars.
    (e) The fraction of cold gas ($f_{\rm cold\ gas} \equiv\ M_{\rm cold\ gas}/(M_{\rm stellar}+M_{\rm cold\ gas})$) as a function of the redshift.
    The gray shade represents the duration of the counter-rotating disk ($V/\sigma>3$ and $\rm cos\,\alpha<0$).} 
    \label{fig:counter_rot}
\end{figure}

During mergers or in the early times of galaxy formation, stars form primarily from unsettled gas.
Following Section~\ref{sec: 4.1}, we identified these stars, which are born \insitu\ with misaligned orbits, using their circularity at birth, and found that they generally contribute to the spheroidal components.
Indeed, many of the stars in spheroid-dominated galaxies at $z=0.7$ have misaligned orbits at birth (leading to lower \fbirthdisk\, in Figure~\ref{fig:m_fbirthdisk}).
However, galaxies sometimes develop counter-rotating disks due to the gas infalling in a direction misaligned with the existing co-rotating disk plane (i.e., sign reversal in the angular momentum of the accreted gas).
This has been found in several numerical studies \citep{Scannapieco2009TheUniverse,Zolotov2015CompactionNuggets,Clauwens2018TheFormation,Garrison-Kimmel2018TheSimulations}, while only a few galaxies have been observed to have such structures \citep[e.g.,][]{Johnston2013Disentangling4550}.
Because the formation of a gaseous disk should precede the development of the second counter-rotating disk, we examined the evolution of the massive galaxies ($>10^{10}\,\msun$) to find such structures.
Of the 24 massive galaxies in our sample, three have developed counter-rotating components.
In our sample, all these three galaxies build counter-rotating structures after mergers.
We show the evolution of one of the galaxies in Figure~\ref{fig:counter_rot}.

Figure~\ref{fig:counter_rot} shows the evolution of one of the galaxies that developed a counter-rotating disk of gas (and/or recent stars).
From the top to the bottom panels, it shows (a) the assembly history of stars \replace{with different origins}, classified as in Section~\ref{sec: 4.1}, (b) the evolution of $D/T$, (c) $V/\sigma$ (gas), (d) the cosine angle between the gas and stellar rotational axes ($\cos\alpha$), and (e) the fraction of cold gas ($f_{\rm cold\ gas} \equiv\ M_{\rm cold\ gas}/(M_{\rm stellar}+M_{\rm cold\ gas})$) as a function of redshift. 
The gas kinematics is also measured using the cold gas ($n_H>10\,{\rm cm^{-3}}$ and $T<2\times10^4\,\rm{K}$) within $R_{90}$. 
To identify a misaligned disk, the $V/\sigma$ of gas was measured with respect to the gas rotational axis (direction of net angular momentum), not with respect to the spin axis of the stars. 
We define that a galaxy develops a counter-rotating structure when $V/\sigma$ of the galactic gas is higher than 3 while the angle ($\alpha$) between the spin axes of gas and stars is greater than $\rm 90^{\circ}$ ($\rm cos\alpha<0$).

As shown in Figure~\ref{fig:counter_rot}, the galaxy encountered a satellite galaxy at $z\sim1.56$ (with a mass ratio of $\sim1:4$), leading to a significant accretion of satellite stars (hatched bars in panel (a)).
The merger ended at $z\sim1.24$, and since then the spin axis of the gas has started to deviate.
While the gas has a high rotational support (high $V/\sigma$ in panel (c)), its spin axis is nearly opposite to the stellar rotational axis ($\cos\alpha\sim-1$ in panel (d)).
Therefore, the stars formed in this counter-rotating disk also have negative circularity at birth (e.g., $\epsilon_{birth}\sim-1$, represented as red bars in panel (a)), thus reducing the fraction of co-rotating disk, $D/T$ (i.e., $f_{\epsilon>0.5}$) in panel (b). 
In panel (b), we also add the mass ratio of the stars with $|\epsilon|>0.5$, co-planar orbits including both co- and counter-rotation, as the green dashed line. 
For most of the time (until $z\sim1.24$), $f_{\epsilon>0.5}$ and $f_{|\epsilon|>0.5}$ do not show much difference, though $f_{|\epsilon|>0.5}$ is systemically higher by definition. 
When there is a sign reversal of the angular momentum of the infalling gas (during the gray shade), however, the counter-rotating disk ($\epsilon<-0.5$) grows, while the existing co-rotating disk ($\epsilon>0.5$) shrinks; therefore the $f_{|\epsilon|>0.5}$ remains almost constant.

This counter-rotating disk lasts more than a Gyr until $z\sim0.85$ when the two spin axes finally align. 
The alignment of the spin axes is caused by the reversal of the stellar rotational axis due to the two combined effects; (i) the kinematics of the original system was dominated by disordered motion (low $D/T$ at $z\sim1.2$); therefore, the total angular momentum of the system was small.  
(ii) Meanwhile, later accretion of gas carries high angular momentum \citep[e.g.,][]{Stewart2011OrbitingAccretion,Kimm2011TheRevisited}; therefore, it re-defines the rotational axis of the system.

\begin{figure*}[h]
    \linespread{1.0}\selectfont{}
    \centerline{\includegraphics[width=\textwidth]{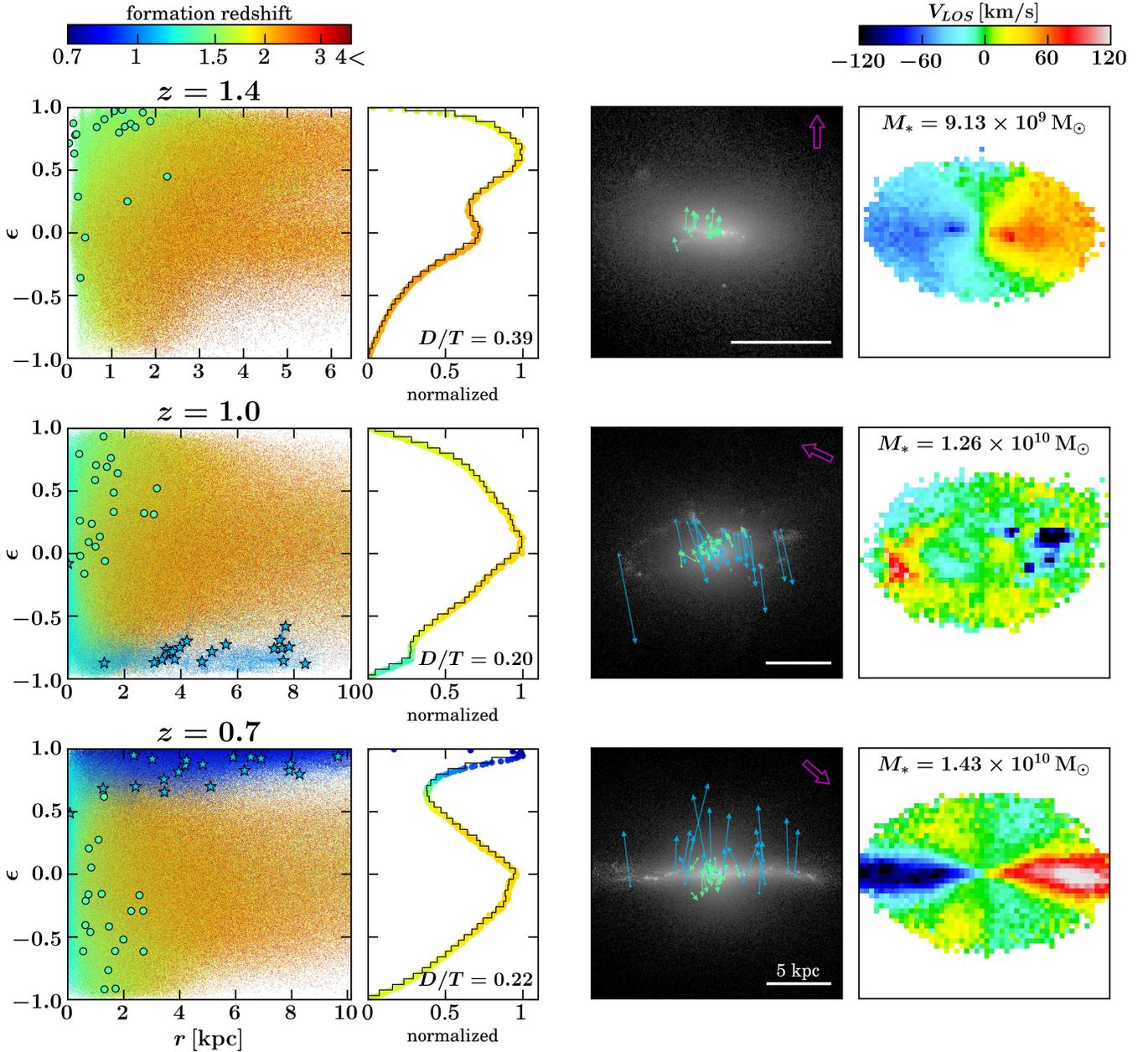}}
    \caption{(First column) Distribution of stars in the plane of circularity and radius for the progenitors of the galaxy tracked in Figure~\ref{fig:counter_rot} at $z=1.4$ (top), $1.0$ (middle), and $0.7$ (bottom). 
    The stars are color-coded according to their formation redshift. 
    As visual guides, 20 randomly selected stars formed at $z=1.4$ are plotted as yellow-green circles and traced on the phase-space diagrams to $z=0.7$. Likewise, 20 random stars formed at $z=1.0$ are traced as light blue stars.
    (Second column) PDFs of the circularity parameters, normalized to the maximum probability and colored according to the average formation redshift.
    (Third column) The edge-on $\sl r$-band images of the progenitors. 
    The \replace{arrows show} the scaled projected angular momentum of the 20 selected stars formed at $z=1.4$ (yellow-green \replace{arrows}) and $z=1.0$ (light blue \replace{arrows}).
    The magenta arrow at the upper right corner of each panel shows the direction of z-axis in a fixed frame of reference (arbitrarily assumed). 
    The white bar represents $\rm 5\,kpc$ (physical scale). 
    (Fourth column) The $\sl r$-band flux weighted line-of-sight velocity map.
    }     
    \label{fig:counter_rot_psd}
\end{figure*}

More explicitly, Figure~\ref{fig:counter_rot_psd} shows how the circularity parameters of the stars change over time as the galaxy develops a counter-rotating disk. 
The first column shows the distribution of the stars of the progenitors of the galaxy at $z=1.4$, $1.0$, and $0.7$ on the circularity versus distance plane.
The color of each star represents its formation redshift.  
As visual guides, we also trace the evolution of 20 randomly selected stars formed at $z=1.4$ (yellow-green circles) and at $z=1.0$ (light blue stars) on the phase-space diagrams.
The PDFs of the circularity parameters of stars in the progenitors, shown in the second column, are colored according to their average formation redshift.
The third and fourth columns show the edge-on $\sl r$-band images of the progenitors and $\sl r$-band flux weighted line-of-sight velocity map. 
The \replace{yellow-green} and \replace{light blue arrows} show the scaled projected angular momentum of the 20 selected stars formed at $z=1.4$ and $z=1.0$, \replace{respectively}.

The stars formed at $z=1.4$ (the yellow-green circles in the top panel) are primarily disk stars at birth (most of them have circularity parameters higher than 0.5).
Therefore, the circularity parameters display a gradual age dependence with younger stars having higher circularity (see the color variation in the right PDF).
Their angular momentum direction is mostly aligned with the galactic rotational axis.
Also, the morphology at $z=1.4$ is fairly disky with $D/T\sim0.4$, as is also seen in the velocity map.

At $z=1.0$ (middle panels), after the merger, most of the stars formed at $z=1.4$ (yellow-green) migrate to the spheroidal component (with lower $\epsilon$). 
Meanwhile, the galaxy produces counter-rotating stars with $\epsilon\sim-1$ (light blue stars), as the star-forming gas rotates along the axis opposite to the galactic spin axis\footnote{Keep in mind that the galactic rotational axis is determined by all the stars within $R_{90}$, the spherically averaged radius containing 90\% of stars in a galaxy.} ($\cos\alpha\sim-1$ in Figure~\ref{fig:counter_rot}(d)).
These young stars start to build a small
peak on the negative side of the circularity PDF (\replace{second column}).
Also, they have much higher angular momentum with opposite direction to the galactic rotational axis (see the light blue arrows at $z=1.0$), compared to the old stars formed at $z=1.4$ (the yellow-green arrows).   
Thus, the $\sl r$-band weighted velocity map shows a counter-rotation in the outer parts due to the young luminous stars. 
Conversely, the central parts, where the preexisting stars are dominant, still exhibit a weak co-rotation.

As the galaxy continues to produce these counter-rotating stars, the galactic rotational axis is reversed and aligned with the rotational axis of the gas.
At $z=0.7$ (bottom panels), the counter-rotating stars formed at $z=1.0$ (light blue stars) have positive circularity, mostly higher than $0.7$, with a significant peak at $\epsilon\sim1$ with young stars (blue) in the PDF.
The angular momentum direction of these stars finally aligned with the galactic rotational axis, and they show clear (co-)rotation in the velocity map.
Interestingly, the stars located apart from the disk plane, which are most likely to be the old stars formed before $z=1.0$, show a mild counter-rotation. 
Even though the young stars contribute significantly to determining the axis of galactic rotation with higher angular momentum, the $D/T$ of the galaxy measured by mass is not sufficiently high to be classified as disk-dominated, because these young stars do not outnumber the preexisitng stars, most of which are spheroid stars.

\section{Discussion}
\label{section5} 

\subsection{Evolution of kinematic morphology}
\label{sec:5.1}
\begin{figure*}
    \linespread{1.0}\selectfont{}
    \includegraphics[width=\textwidth]{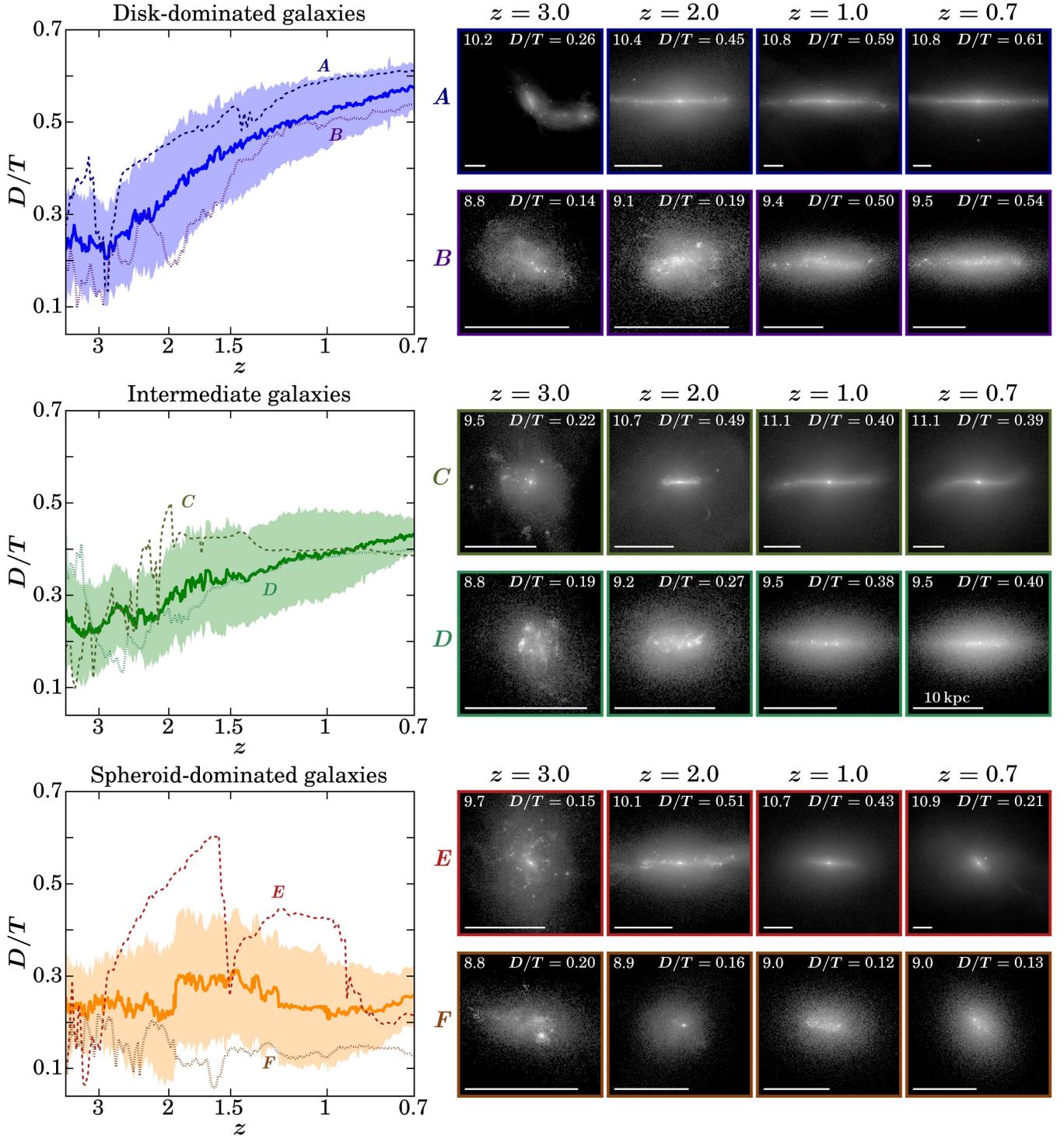}
    \caption{(Left) $D/T$ evolution of galaxies with different morphologies. The solid lines indicate the average $D/T$ of the disk-dominated (blue line), \replace{intermediate (green line)}, and spheroid-dominated (\replace{orange} line) galaxies, and the shading represents 1$\sigma$. 
    We also plot the $D/T$ evolution of individual galaxies as specimens of massive disk-dominated
    (\qsay{$A$}, dashed dark blue line), low-mass disk-dominated (\qsay{$B$}, dotted purple line), \replace{massive intermediate (\qsay{$C$}, dashed dark green line), low-mass intermediate (\qsay{$D$}, dotted green line)}, massive spheroid-dominated (\qsay{$E$}, dashed dark red line), and low-mass spheroid-dominated (\qsay{$F$}, dotted brown line) galaxies.
    (Right) The $\sl r$-band (in rest frame) edge-on images of those specimens from $z=3.0$ to $z=0.7$.} 
    \label{fig:morp_z}
\end{figure*}

In this section, we explore the evolution of galaxies with different kinematic morphology (final disk-to-total ratio).
We selected disk-dominated, \replace{intermediate} and spheroid-dominated galaxies at $z=0.7$ and tracked back their kinematic morphological evolution to $z=4$, which is shown in the left panels of Figure~\ref{fig:morp_z}. 
In addition to the average growth of $D/T$ in \replace{the three} groups, we also added the $D/T$ evolution of individual galaxies as specimens of \replace{massive and low mass disk-dominated (``$A$'',``$B$''), intermediate (``$C$'',``$D$'') and spheroid-dominated galaxies (``$E$'',``$F$'').}  
The right panels of Figure~\ref{fig:morp_z} show the $\sl r$-band edge-on images of those specimens at $z=3.0$, $2.0$, $1.0$, and $0.7$.

At $z\sim3$, the disk component is still negligible in all of our galaxies, resulting in $D/T$ between 0.2 and 0.3. 
Disk-dominated galaxies (selected at $z=0.7$), however, keep growing their disks until $z=0.7$. 
The increase in the disk component is more pronounced in massive galaxies than in low-mass ones. 
As illustrated in Figure~\ref{fig:morp_z}, the massive galaxy (\qsay{$A$}) has a well-developed disk from $z\sim2$, while the low-mass galaxy (\qsay{$B$}) has one from approximately $z\sim1$.

Low-mass spheroid-dominated galaxies (\replace{\qsay{$F$}}), conversely, have not been able to predominantly form disk stars and, therefore, have continually had low $D/T$.
In our sample, the massive spheroid-dominated galaxies (\replace{\qsay{$E$}} as an example) seem to be formed via severe disk disruption (Figure~\ref{fig:sph_growth}(c)) and high accretion (Figure~\ref{fig:m_fbirthdisk}(a)); therefore, some of them once had well-developed disks with peaks in the $D/T$ evolution. 
Indeed, as a visual inspection in the right panel shows, 
the sample galaxy \qsay{\replace{$E$}} developed a disk at $z\sim2$, only to be disrupted and become a spheroid-dominated galaxy at $z=0.7$.
Some galaxies develop counter-rotating disks, as mentioned in Section~\ref{sec:4.4}. 
Unfortunately, we lack massive spheroids in the sample because our galaxies are confined to a field environment where dramatic merger-driven morphological transformations are rare.

\subsection{Contribution from different channels to disk and spheroidal components}
\label{sec:5.2}
\begin{figure*}[h]
    \centering
    \linespread{1.0}\selectfont{}
    \includegraphics[width=\textwidth]{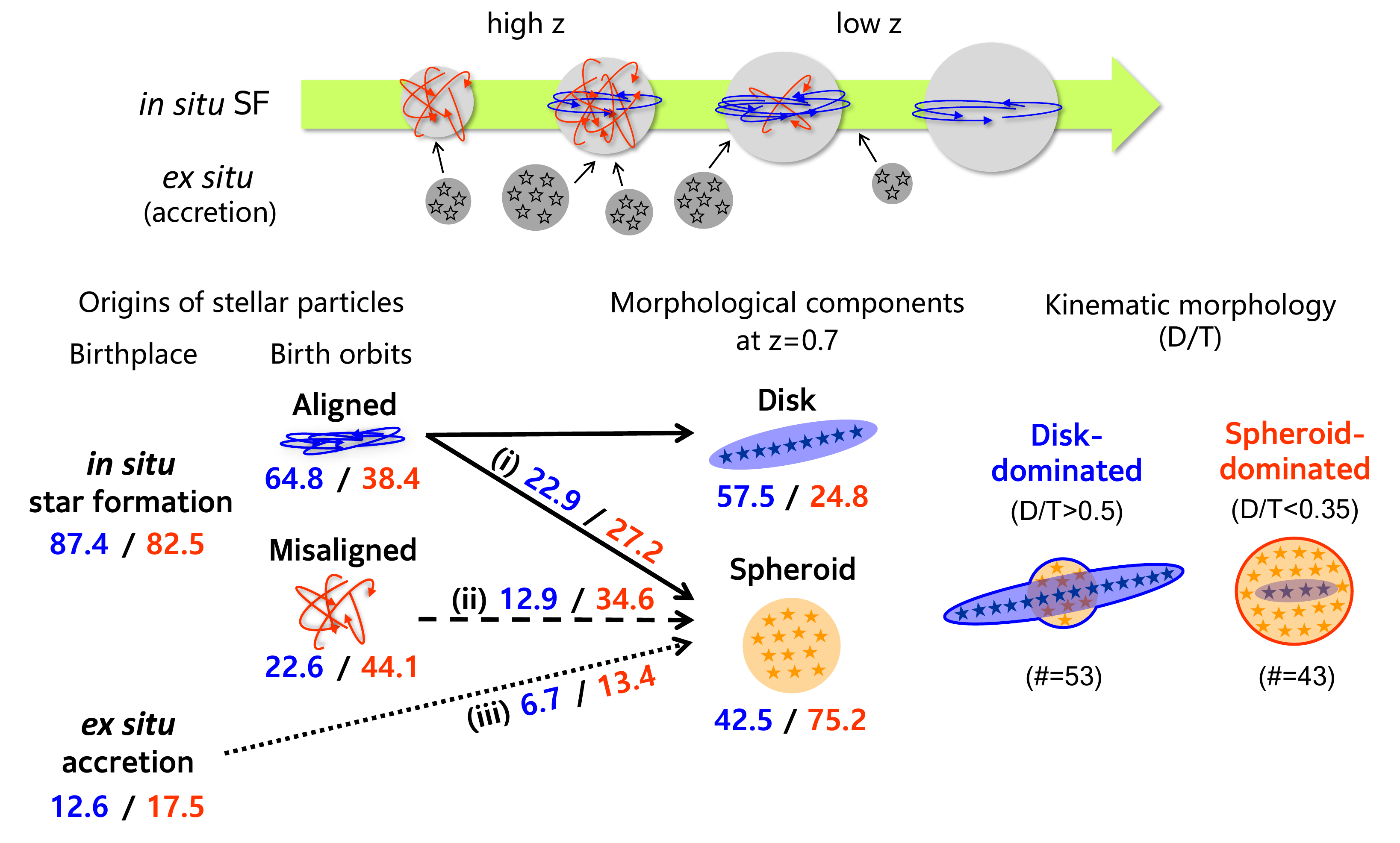}
    \caption{(Top) A general picture of the evolution of a galaxy based on the discussion of Section~\ref{sec: 4.2}. Galaxies grow from \insitu\, star formation and accretion of stars that are formed \exsitu. As redshift decreases, stars are more likely to form with aligned orbits (blue orbits).
    (Bottom) A Schematic diagram showing the estimates (in percentage) of stellar particles with different origins in the disk and spheroidal components (same as Figure~\ref{fig:intro}). Disk and spheroidal components of galaxies are kinematically decomposed at $z=0.7$, and the arrows indicate the different channels to the spheroids: (i) stars initially formed in the disk (``aligned'' initial orbits) yet migrated to the spheroid, (ii) stars born with ``misaligned'' initial orbits, and (iii) accreted stars. The numbers shown in blue and orange colors are the average estimates for the 53 disk-dominated galaxies ($D/T>0.5$) and the 43 spheroid-dominated galaxies ($D/T<0.35$), respectively.}
    \label{fig:conclusion}
\end{figure*}

The top panel of Figure~\ref{fig:conclusion} presents a general picture of the evolution of galaxies from an extrapolation based on Section~\ref{section4}; galaxies grow via \insitu\, star formation and accretion (of stars that are formed \exsitu). In the early stage of galaxy formation, stars mostly form with misaligned orbits (orange) through numerous mergers between proto-galaxies. As redshift decreases, most stars are formed with aligned orbits (blue), though the star formation rate is decreasing.
As introduced in Figure~\ref{fig:intro}, we divide the possible channels that contribute to the disk and spheroidal components of a galaxy.
Let us quantify the importance of these channels and revisit the different kinematic morphology of galaxies (probed by disk-to-total ratio).
The bottom of Figure~\ref{fig:conclusion} provides the estimates (in percentage) of stars with different origins in the galaxies.
The numbers shown in blue and orange colors are the average estimates for the 53 disk-dominated galaxies ($D/T>0.5$) and the 43 spheroid-dominated galaxies ($D/T<0.35$) selected at $z=0.7$.
Table~\ref{table:6.1} also presents the estimates averaged over the massive ($>10^{10}\,\msun$) and the low-mass ($10^9$--$10^{10}\,\msun$) galaxies and the associated standard deviation.
Based on the estimates, we summarize the formation of galaxies with different kinematic morphology (disk-to-total ratio) as follows.

Disk-dominated galaxies: 
approximately 90\% of the stars in the disk-dominated galaxies are formed \insitu, and many of them (64.8\% in total, \replace{\fbirthdisk$\sim0.74$}) are born with aligned orbits with respect to the rotating plane.
We also found that there is a mass dependency for galaxies to start disk-mode star formation (Sections~\ref{sec: 4.2}).
Therefore, higher fractions of stars are born as disk stars in heavier galaxies (\fbirthdisk$\sim0.82$) compared to lower-mass galaxies (\fbirthdisk$\sim0.71$).
Some of the disk-born stars (22.9\% in total, solid arrow) become part of the spheroids as their orbits are perturbed; therefore, approximately half of the spheroidal components originate from the disks (Section~\ref{sec:4.3}).
Finally, 12.6\% of the stars in disk-dominated galaxies are formed \exsitu\ and around a half of them (or 6.7\% in total, dotted arrow) contribute to the growth of the spheroids.

Spheroid-dominated galaxies:
unlike disk-dominated galaxies, many of the stars in the spheroid-dominated galaxies are born with misaligned orbits (44.1\% in total, \fbirthdisk$\sim0.47$) and contribute to the spheroidal components (dashed arrow).
Massive galaxies have a higher fraction of disk stars at birth (\fbirthdisk$\sim0.61$) than low-mass galaxies (\fbirthdisk$\sim0.45$); however, most of \replace{the disk stars} eventually migrate to the spheroidal components (see Figure~\ref{fig:sph_growth}(c)).
Indeed, the transformation from disks to spheroids ((i), solid arrow) is the most important channel (30.8\% in total) for growing the spheroidal components in massive galaxies.
On average, spheroid-dominated galaxies have higher accretion levels (17.5\% in total) than disk-dominated galaxies, a trend which is more pronounced for massive galaxies (see also Figure~\ref{fig:m_fbirthdisk}(a)). 
In addition, a higher fraction of \exsitu\ stars contributes to the spheroidal components.

\begin{table*}[b]
\centering
\caption{Stellar origins of the galaxies with different mass and morphology}
\label{table:6.1}

\begin{tabular}{p{2cm}||p{3cm}|p{5cm}|p{0.15\linewidth}||p{0.15\linewidth}|p{0.15\linewidth}|p{0.15\linewidth}|p{0.15\linewidth}}
\hline
\multicolumn{1}{c}{\multirow{2}{*}{}} & \multicolumn{3}{c}{\multirow{2}{*}{Origins of stellar particles}} & \multicolumn{4}{c}{\multirow{2}{*}{Morphological components}} 
\\
\multicolumn{1}{c}{} & \multicolumn{3}{c}{} & \multicolumn{4}{c}{} 
\\
\hline
\multicolumn{1}{c}{\multirow{2}{*}{}} & 
\multicolumn{2}{c}{\insitu} & \multicolumn{1}{c}{\multirow{2}{*}{\exsitu}} & 
\multicolumn{1}{c}{\multirow{2}{*}{Disk}} & 
\multicolumn{3}{c}{Spheroid (i+ii+iii)}    \\ 
\multicolumn{1}{c}{} & \multicolumn{1}{c}{Aligned (\fbirthdisk)} & \multicolumn{1}{c}{Misaligned} & \multicolumn{1}{c}{} & \multicolumn{1}{c}{} & \multicolumn{1}{c}{(i)} & \multicolumn{1}{c}{(ii)} & \multicolumn{1}{c}{(iii)} 
\\ 
\hline
\hline
\multicolumn{1}{c}{\bf Disk-dominated (53)} & 
\multicolumn{1}{c}{64.8$\pm$9.1 (\replace{0.74})} & \multicolumn{1}{c}{22.6$\pm$7.1} & \multicolumn{1}{c}{12.6$\pm$8.6} & \multicolumn{1}{c}{57.5$\pm$5.2} & \multicolumn{1}{c}{22.9$\pm$4.4} & \multicolumn{1}{c}{12.9$\pm$4.2} & \multicolumn{1}{c}{6.7$\pm$4.7} 
\\ 
\hline
\multicolumn{1}{c}{Massive disk (15)} 
& \multicolumn{1}{c}{66.5$\pm$7.4 (0.82)} & \multicolumn{1}{c}{14.8$\pm$4.3} & \multicolumn{1}{c}{18.7$\pm$8.2} & \multicolumn{1}{c}{59.3$\pm$5.2} & \multicolumn{1}{c}{19.9$\pm$4.3} & \multicolumn{1}{c}{9.4$\pm$2.7} & \multicolumn{1}{c}{11.4$\pm$5.2} 
\\ 
\hline
\multicolumn{1}{c}{Low-mass disk (38)} 
& \multicolumn{1}{c}{64.1$\pm$9.6 (0.71)} & \multicolumn{1}{c}{25.7$\pm$5.4} & \multicolumn{1}{c}{10.2$\pm$7.5} & \multicolumn{1}{c}{56.8$\pm$5.0} & \multicolumn{1}{c}{24.1$\pm$3.8} & \multicolumn{1}{c}{14.3$\pm$3.9} & \multicolumn{1}{c}{4.9$\pm$2.8} 
\\ 
\hline
\hline
\multicolumn{1}{c}{\bf Spheroid-dominated (43)} & 
\multicolumn{1}{c}{38.4$\pm$10.8 (0.47)} & \multicolumn{1}{c}{44.1$\pm$13.4} & \multicolumn{1}{c}{17.5$\pm$12.4} & \multicolumn{1}{c}{24.8$\pm$6.6} & \multicolumn{1}{c}{27.2$\pm$8.1} & \multicolumn{1}{c}{34.6$\pm$11.4} & \multicolumn{1}{c}{13.4$\pm$9.2} 
\\ 
\hline
\multicolumn{1}{c}{Massive spheroid (5)} & 
\multicolumn{1}{c}{41.1$\pm$8.6 (0.61)} & \multicolumn{1}{c}{26.5$\pm$8.0} & \multicolumn{1}{c}{32.4$\pm$7.8} & \multicolumn{1}{c}{27.8$\pm$5.5} & \multicolumn{1}{c}{30.8$\pm$11.0} & \multicolumn{1}{c}{19.8$\pm$4.8} & \multicolumn{1}{c}{21.7$\pm$5.7} 
\\ 
\hline
\multicolumn{1}{c}{Low-mass spheroid (38)} & 
\multicolumn{1}{c}{38.0$\pm$11.0 (0.45)} & \multicolumn{1}{c}{46.4$\pm$12.1} & \multicolumn{1}{c}{15.5$\pm$11.5} & \multicolumn{1}{c}{24.4$\pm$6.6} & \multicolumn{1}{c}{26.7$\pm$7.5} & \multicolumn{1}{c}{36.6$\pm$10.5} & \multicolumn{1}{c}{12.3$\pm$9.0} 
\\
\hline
\hline
\multicolumn{1}{c}{\bf All (144)} & 
\multicolumn{1}{c}{54.2$\pm$14.4 (0.62)} & \multicolumn{1}{c}{32.4$\pm$13.0} & \multicolumn{1}{c}{13.3$\pm$10.4} & \multicolumn{1}{c}{42.8$\pm$14.3} & \multicolumn{1}{c}{26.0$\pm$6.6} & \multicolumn{1}{c}{22.2$\pm$11.5} & \multicolumn{1}{c}{9.0$\pm$7.6} 
\\
\hline

\end{tabular}
\begin{tablenotes}
\small
\item Note. --The estimates (in percentage) of the origins of stellar particles (\insitu\, and \exsitu) \replace{in the galaxies and in the morphological components, disk and spheroid (see Figure~\ref{fig:conclusion})}. 
\replace{Each row presents} the estimates averaged for disk-/spheroid-dominated galaxies (53 and 43 galaxies, respectively) and all the galaxies with masses greater than $10^9\,\msun$ (144 galaxies, \replace{selected at $z=0.7$}) \replace{and the associated} one standard deviation. Note that all these quantitative estimates are highly sensitive to the definition of disk and spheroidal component (also, \replace{the definition of} aligned and misaligned initial orbits).
\end{tablenotes}
\end{table*}

\subsection{Caveats}
\label{sec:5.3}

Throughout this study, we used a simple circularity cutoff of $\epsilon=0.5$ (or $\epsilon_{birth}=0.5$) for identifying disk (or disk-born) stars. 
In addition, our study suffers from the fact that the New Horizon simulation has ended at $z=0.7$, covering only half of the cosmic history. 
While there is no absolute criterion to decompose a galaxy into disk and spheroid, we discuss some limitations that may arise from applying this simple cutoff and exploring only the field galaxies down to $z=0.7$.

In New Horizon at $z=0.7$, even the galaxy with the highest value of $D/T$ (which is $\sim0.7$, see Figure~\ref{fig:m_dt}) has a considerable spheroidal component ($\sim0.3$). 
The absence of pure kinematic disk galaxies in our sample partially arises from the shortcoming of our kinematic decomposition method\replace{; we} consider all the stars
with $\epsilon<0.5$ to be part of the spheroid. 
This means that even if all the stars of a galaxy followed a single distribution function, strongly peaked at $\epsilon=1.0$, its $\epsilon<0.5$ tail would be counted as part of a spheroidal component.
Similarly, disk components also have contamination from spheroids especially in \replace{the} spheroid-dominated galaxies. 
If we assume a Gaussian distribution of the circularity parameter for the bulk of randomly-orbiting stars, the stars on the right tail above $\epsilon>0.5$ are classified as disk stars.

The fact that our sample is biased to the field environment and explored only the first half of the cosmic time is another issue we should address. We have discussed in Section~\ref{sec:3.2} that this might lead to the discrepancy in the morphological mix between the New Horizon and the local galaxies: the New Horizon lacks massive elliptical galaxies. In the more dense environment like clusters, the majority of massive galaxies are early-type even at $z\sim0.7$ \citep[e.g.,][]{Smith2005EvolutionGalaxies,vanderWel2007TheGalaxies}. From a quantitative point of view, the importance of \exsitu\, (accreted) stars (described as the dotted arrow (iii) in Figure~\ref{fig:conclusion}) in cluster galaxies would be heightened compared to the New Horizon sample.

To sum up, the quantitative description of the channels discussed in Section~\ref{sec:5.2} is, therefore, highly sensitive to the definition of the disk and spheroid. 
Nonetheless, we assumed that the consistent cutoff of $\epsilon=0.5$ is reasonable enough to disentangle the disk and spheroidal components of high-redshift galaxies undergoing dramatic evolution.
Given that the vast majority of galaxies in the local Universe are located in the field, the conclusions drawn from our sample should still hold for the bulk of the galaxy population.
While the estimates we provide may not be suitable for direct comparison with the local universe, we expect these to be the guiding values for those from future high-resolution large-volume simulation down to $z=0$ and observational studies.

\section{Conclusions}
\label{section6} 

Using the New Horizon simulation, we explored the origin of disk and spheroidal components of 144 field galaxies with masses greater than $10^9\,\msun$. 
We decomposed the simulated galaxies into disks and spheroids based on the orbital properties of their stellar particles.
We traced the origins of stellar particles according to where they formed and the properties of their orbits at birth (see Section~\ref{sec: 4.1}). 
Our main results can be summarized as follows.

\begin{itemize}
  \item Galaxies form disk stars in a mass dependent way; massive galaxies ($\it M_{*,\rm z=0.7}\rm >10^{10}\,\msun$) start to develop their disks at $z\sim1-2$, while low-mass galaxies ($\rm 10^{9}\,\msun< \it M_{*,\rm z=0.7}\rm <10^{10}\,\msun$) do at $z<1$ (Section~\ref{sec: 4.2} and see also Dubois et al. in prep.). 
  \replace{Indeed, the fraction of disk stars at birth (\fbirthdisk) increases with stellar mass; most of the stars in the massive galaxies formed in the disks (\fbirthdisk$\sim$0.76), whereas a half of the low-mass galaxies in our sample have not yet developed their disks by $z\sim0.7$.} 
  \replace{The formation of disks is affected by accretion as well; galaxies with higher accretion have lower \fbirthdisk, as mergers or fly-bys can boost gas turbulence in galaxies, making galaxies less likely to form co-rotating disk stars.} 
  \replace{Therefore,} the fraction of disk stars at birth depends on both mass and accretion history (Figure~\ref{fig:m_fbirthdisk}).    
  
  \item Stars with disk-origin contribute significantly to the spheroidal components except for the low-mass spheroid-dominated population, as they have not formed much disk stars by $z\sim0.7$ \replace{(Figure~\ref{fig:sph_growth})}. 
  In the disk-dominated galaxies (both massive and low-mass), these stars with disk-origin make up around half of their spheroidal component. 
  Accretion plays an important role in the spheroidal components in the massive galaxies, especially massive spheroid-dominated galaxies (accounting for $\sim30\%$ of the spheroid stars).
  
  \item The development of counter-rotating disk component is not very rare in the evolution of massive galaxies; three of 24 massive galaxies ($\sim12.5\%$) in our sample have had such structures.
  These counter-rotating structures can last for more than a Gyr, until they become the dominant component, and flip the angular momentum of the galaxy in the opposite direction. 
  As a result, kinematically decoupled features (with old star, in the inner regions, counter-rotating while young stars formed recently co-rotating) appear on (edge-on) velocity maps (Figure~\ref{fig:counter_rot_psd}).
 
\end{itemize}

In conclusion, disk and spheroidal components of galaxies are formed via several processes, and the kinematic morphology of galaxies (i.e., disk-to-total ratio) is determined by the significance of each process.
Because the importance of the processes (e.g., disk formation, migration of disk stars to spheroidal components, and accretion of disrupted satellite stars) depends on both {\em stellar mass} and {\em accretion history}, the morphology of galaxies also needs to be understood in relation to them.

Identifying the origin and kinematics of individual stars in a galaxy is extremely challenging in observations.   
Only in the Milky Way Galaxy has it been possible to obtain the kinematic information of individual stars, e.g., via RAVE \citep{Steinmetz2006TheRelease}, \textit{Gaia} \citep{GaiaCollaboration2016GaiaService} and APOGEE \citep{Majewski2017TheAPOGEE}.
The kinematic properties of external galaxies are primarily studied with IFU surveys \citep{Emsellem2007TheGalaxies,Cappellari2007TheKinematics,Croom2012TheSpectrograph,Sanchez2012CALIFAPresentation,Bundy2015OverviewObservatory}, more actively at lower redshifts.
We expect that future observations will provide more detailed information on the origins of morphological components in galaxies at high redshifts, too.

\bigskip 

\acknowledgements
We thank the referee for constructive comments. S.K.Y, the corresponding author, acknowledges support from the Korean National Research Foundation (NRF-2017R1A2A05001116). 
This research was supported in part by the National Science Foundation under Grant No. NSF PHY-1748958. 
This work was granted access to the high-performance computing resources of CINES under the allocations c2016047637 and A0020407637 from GENCI, and KISTI (KSC-2017-G2-0003). 
Large data transfer was supported by KREONET, which is managed and operated by KISTI. 
This work relied on the HPC resources of the Horizon Cluster hosted by Institut d’Astrophysique de Paris. 
We warmly thank S. Rouberol for running the cluster on which the simulation was post-processed.
T.K. was supported by the Yonsei University Future-leading Research Initiative (RMS2-2018-22-0183). 
H.C. acknowledges the support by Norwegian Research Council Young Research Talents Grant 276043 “Simulating the Circumgalactic Medium and the Cycle of Baryons In and Out of Galaxies Throughout Cosmic History”
The research of J.D. is supported by Adrian Beecroft and STFC.
M.V. acknowledges funding from the European Research Council under the European Community's Seventh Framework Programme (FP7/2007-2013 Grant Agreement no. 614199, project “BLACK"). 
S.K. acknowledges a Senior Research Fellowship from Worcester College Oxford. This work is partially supported by the Spin(e) grant ANR-13- BS05-0005 of the French Agence Nationale de la Recherche, and ERC grant 670193.

\clearpage
\bibliography{reference}

\end{document}